\documentclass[%
superscriptaddress,
nofootinbib,
 amsmath,amssymb,
 aps,
]{revtex4-2}

\usepackage{graphicx}
\usepackage{dcolumn}
\usepackage{bm}
\usepackage{latexsym}
\usepackage{comment}
\usepackage{amssymb}
\usepackage{multirow}
\usepackage{ulem}
\usepackage{lineno}
\usepackage{soul,graphicx,xcolor}
\usepackage{slashed}
\usepackage{bm}

\definecolor{webgreen}{rgb}{0,0.75,0}
\definecolor{webred}{rgb}{0.75,0,0}
\definecolor{webblue}{rgb}{0,0,0.75}
\definecolor{darkblue}{rgb}{0,0,0.6}
\definecolor{dunkelgrau}{rgb}{0.8,0.8,0.8}
\definecolor{lgray}{rgb}{0.95,0.95,0.95}
\definecolor{lgreen}{rgb}{0.95,1.00,0.90}
\definecolor{dgreen}{rgb}{0.1,0.75,0.1}
\definecolor{lblue}{rgb}{0.9,0.95,1.00}
\definecolor{lred}{rgb}{1.00,0.92,0.85}
\definecolor{shadecolor}{rgb}{1.00,0.92,0.82}
\usepackage[colorlinks=true,linkcolor=darkblue ,citecolor=dgreen,urlcolor=darkblue]{hyperref}

\newcommand{\qqbar}{$\rm q\bar{\rm q}$}

\newcommand{\auau}{$\textrm{Au}$-$\textrm{Au}$}

\newcommand{\agev}{$A$GeV}

\newcommand{\gev}{GeV}
\newcommand{\mev}{MeV}
\newcommand{\fmc}{fm/$c$}

\newcommand{\eg}{{\it e.g.}}
\newcommand{\ie}{{\it i.e.}}

\newcommand{\ee}{$\rm e^+ \rm e^-$}

\newcommand{\lsim}{\lesssim}

\newcommand{\beq}{\begin{equation}}
\newcommand{\eeq}{\end{equation}}
\newcommand{\bea}{\begin{eqnarray}}
\newcommand{\eea}{\end{eqnarray}}
\newcommand{\bef}{\begin{figure}}
\newcommand{\eef}{\end{figure}}
\newcommand{\bce}{\begin{center}}
\newcommand{\ece}{\end{center}}
\begin{document}
%
\preprint{APS/123-QED}
\title{Dilepton signature of a first-order phase transition}
%
\author{Florian Seck}
\affiliation{Technische Universit\"at Darmstadt, 64289 Darmstadt, Germany}
\author{Tetyana Galatyuk}
\affiliation{GSI Helmholtzzentrum f\"ur Schwerionenforschung GmbH, 64291 Darmstadt, Germany}
\affiliation{Technische Universit\"at Darmstadt, 64289 Darmstadt, Germany}
\author{Ayon Mukherjee}
\affiliation{ E\"otv\"os Lor\"and Tudom\'anyegyetem (ELTE), 1053~Budapest, Hungary}
\author{ Ralf Rapp}
\affiliation{ Texas A$\&$M University, College Station, Texas 77843-3366, USA}
\author{Jan Steinheimer}
\affiliation{Frankfurt Institute for Advanced Studies, 60438~Frankfurt, Germany}
\author{Joachim Stroth}
\affiliation{Goethe-Universit\"{a}t, 60438~Frankfurt, Germany}
\affiliation{GSI Helmholtzzentrum f\"ur Schwerionenforschung GmbH, 64291 Darmstadt, Germany}
\affiliation{Helmholtz Research Academy Hesse for FAIR, Campus Frankfurt, 60438 Frankfurt, Germany}
\author{Maximilian Wiest}
\affiliation{Technische Universit\"at Darmstadt, 64289 Darmstadt, Germany}
\date{\today}
%
\begin{abstract}
The search for a first-order phase transition in strongly interacting matter is one of the major objectives in the exploration of the phase diagram of quantum chromodynamics (QCD).
In the present work we investigate dilepton radiation from the hot and dense fireballs created in Au-Au collisions at projectile energies of $1-2$\,$A$GeV for potential signatures of a first-order transition.
Toward this end, we employ a hydrodynamic simulation with two different equations of state, with and without a phase transition.
The latter is constrained by susceptibilities at vanishing chemical potential from lattice-QCD as well as neutron star properties, while the former is implemented via modification of the mean fields in the quark phase.
We find that the latent heat involved in the first-order transition leads to a substantial increase in the low-mass thermal emission signal by about a factor of two above the crossover scenario. 
\end{abstract}
\keywords{Dileptons, hydro, phase transition}
\maketitle
%
%
\section{Introduction}
\label{intro}
%
Mapping out the phase structure of strong-interaction matter under extreme conditions of temperature and density is a central goal of (ultra-) relativistic heavy-ion collisions (HICs). 
Numerical evaluations of quantum chromodynamics (QCD) on a discrete space-time lattice, so-called lattice QCD (lQCD), have revealed that the early universe, where the baryon chemical potential $\mu_{\rm{B}}$ was very small, evolved via a smooth crossover transition from a state of deconfined quarks and gluons into a medium of color-neutral hadrons with spontaneously broken chiral symmetry~\cite{Bazavov:2019lgz}. 
The observations made to date in ultrarelativistic HIC experiments at 
the Super Proton Synchroton (SPS), Relativistic Heavy-Ion Collider (RHIC), and the Large Hadron Collider (LHC) generally support the lQCD result.
In particular, a chemical freeze-out temperature was extracted from the composition of the observed hadron species~\cite{Andronic:2017pug} which coincides, within uncertainties, with the pseudo-critical temperature of $T_{\rm pc}\simeq 155$ MeV extracted from lQCD at $\mu_{\rm{B}}=0$~\cite{Borsanyi:2013bia,Bazavov:2018mes}. 

Contrary to the investigation of QCD matter at small baryon chemical potential in HICs at the high-energy frontier, the structure of the phase diagram at high baryochemical potentials is much less explored.
This region can be accessed experimentally by lowering the collision energy.
Due to the initial energy degradation of the nucleons, lower collision energies produce a 
central zone with increasing net baryon density (\ie, higher $\mu_{\rm{B}}$, accompanied by a reduced production of antibaryons), until the matter compression decreases again at still lower energies.
As lQCD computations become rather challenging with rising $\mu_{\rm{B}}$, theoretical guidance in this region of the phase diagram largely 
resorts to effective theories or QCD-inspired models. 
Many of these approaches predict the occurrence of a first-order transition between hadronic and deconfined matter, often embedded into a rich structure of hitherto unknown phases of QCD matter.

Several new experimental facilities are aimed at exploring the high-baryon density region of the phase diagram~\cite{Galatyuk:2019lcf}, and various observables have been suggested as key facilitators in this effort. 
For example, the azimuthal-angular dependence of the transverse-momentum spectra of hadrons, associated with collective flow patterns developed by the medium in noncentral 
collisions~\cite{Brachmann:1999xt,Nara:2016phs,Nara:2018ijw,Spieles:2020zaa}, has been proposed as a signal of 
a softening of the equation of state (EoS), most notably within fluid 
dynamical simulations of the expanding fireball. However, these simulations usually depend on 
parameters that specify the initial-state configurations and freeze-out criteria~\cite{Petersen:2009mz,Steinheimer:2009nn}, which significantly 
affect their interpretations; in addition hadronic rescatterings in the more dilute phases 
tend to wash out the information from the EoS-driven expansion~\cite{Steinheimer:2017vju}. Alternatively, 
the formation of density fluctuations caused by a spinodal decomposition or nucleation in a first-order phase transition~\cite{Steinheimer:2012gc,Herold:2013qda}, or fluctuations at a second-order critical endpoint (see, \eg, Refs.~\cite{Nahrgang:2011mg,Herold:2013bi,Stephanov:2017ghc,Bzdak:2019pkr,Bluhm:2020mpc}), have been widely investigated. 
However, it remains a challenging question whether fluctuations in coordinate space can be preserved in hadronic final-state observables when using realistic simulations of the fireball~\cite{Steinheimer:2019iso,Spieles:2020zaa}.

Electromagnetic (EM) radiation does not suffer from several of the aforementioned disadvantages.
As photons and dileptons are emitted continuously throughout the fireball evolution with negligible final-state interactions, they provide a unique tool to probe the microscopic properties of QCD matter created in (ultra-) relativistic HICs~\cite{Rapp:2009yu}.
Over the past two decades, it has been found that the measured dilepton (and with some caveat photon~\cite{Gale:2020xlg}) spectra at the SPS, RHIC, and the LHC can be successfully described using thermal emission rates from the QGP (calculated, \eg, in hard-thermal-loop perturbation theory and/or constrained by lQCD) and hadronic matter (calculated using hadronic many-body theory).
In particular, the melting of the $\rho$ meson causes the hadronic rates to merge into the QGP ones in the expected transition region, which indicates a change in degrees of freedom from hadrons to partons and is also compatible with chiral symmetry restoration~\cite{Hohler:2013eba}. 
These findings are rather robust against details of the space-time evolution.
For example, time-dependent thermal blast-wave expansions~\cite{Rapp:1999ej,vanHees:2007th}, relativistic fluid dynamics~\cite{Vujanovic:2013jpa}, and microscopic transport simulations \cite{Endres:2014zua,Galatyuk:2015pkq} give rather similar results as long as the final-state hadron yields and spectra are consistent with the measured ones. 
As such, the current interpretation of dilepton and photon data does not exhibit indications of a first-order transition. 
However, this situation may change at lower collision energies.
For example, it has recently been suggested that EM radiation may be sensitive to a local density clumping and a longer lifetime of the fireball when the medium undergoes a first-order phase transition~\cite{Rapp:2014hha,Li:2016uvu}, see also Ref.~\cite{Cleymans:1986na} for an early investigation of dilepton transverse-momentum spectra for different phase transition scenarios.

The purpose of the present work is to investigate whether dilepton radiation in nuclear collisions at beam energies of the Schwer-Ionen-Synchroton (SIS18) facility are sensitive to the presence of a first-order phase transition.
In particular, the associated latent heat is generally expected to lead to a significantly different bulk evolution compared to a crossover transition, and we will quantify how this difference manifests itself in the thermal dilepton emission from both scenarios.
In addition to two ideal-hydrodynamic evolutions, we will also compare the results to our previous calculations using coarse-grained transport simulations of the fireball in 1.23\,\agev\ Au-Au collisions~\cite{Galatyuk:2015pkq}.
This will allow us to obtain an estimate of the uncertainty in the space-time evolution by using both the small and large mean-free-path
limits (hydrodynamics and kinetic transport, respectively) for the background medium in the evaluation of dilepton emission, while the effects of viscosities on the fluid-dynamical evolution will be deferred to a future study.

Our article is organized as follows. In Sec.~\ref{sec:emissivity} we introduce the dilepton emissivity of hot and dense matter, how it relates to the EM spectral function of the medium, and how it is employed to compute dilepton spectra in heavy-ion collisions. 
In Sec.~\ref{sec:bulk}, we first recall our previously used coarse-graining procedure of the ultra-relativistic quantum molecular dynamics (UrQMD) transport model (Sec.~\ref{ssec:urqmd}) and then lay out the basic ingredients to our hydrodynamic evolution model (Sec.~\ref{ssec:fluid}), including a discussion of the two equations of states used in this work (Sec.~\ref{ssec:eos}). 
In Sec.~\ref{sec:results} we benchmark the final-state pion spectra of the different evolution models against each other (Sec.~\ref{ssec:pions}), inspect the space-time evolutions in terms of temperature and baryon density profiles (Sec.~\ref{ssec:Tmu-evo}), and present and discuss the pertinent dilepton spectra (Sec.~\ref{ssec:dl-spec}). We also address the occurrence of (presumably artificial) high-temperature hydro cells at the fringes of the fireball in the hydro evolution and their potential impact on our results. 
We summarize and conclude in Sec.~\ref{sec:concl}.
%
\section{Dilepton Emissivity and Spectra}
\label{sec:emissivity}
%
The phase-space distributions of dileptons provide valuable insights into the temperature, collectivity, and spectral structure of the medium created during a nuclear collision.
Generally speaking, large values of the invariant mass, $M$, and three-momentum, $q$, of the lepton pair are associated with early emission in the evolution history, although momentum distributions are also blue-shifted by the collective flow of the medium which becomes large(r) in the later stages.

Our starting point is the differential emissivity of the medium,
\begin{equation}
    \varepsilon = \frac{{\rm{d}}N_{\rm ll}}{\rm{d}\it V \rm{d} \it t \rm{d}^4 \it q} \ , 
\label{emissivity}
\end{equation}
which defines the radiation rate of virtual photons from a cell of strongly interacting matter per unit time and four-momentum.
In thermal equilibrium~\cite{Pisarski:1981mq,McLerran:1984ay}, one has
\begin{equation}
    \varepsilon = -\frac{\alpha^2_{\rm EM}}{\pi^3} \frac{ {\rm {L}}(M)}{M^2} f^B(q_0;T)~{\rm Im}\Pi_{\rm EM}(M,q;\mu_{\rm {B}},T)  
\label{emisthermeq}
\end{equation}
where $\alpha_{\rm EM}$ is the EM coupling constant, ${\rm Im}{\Pi_{\rm EM}}$ the EM spectral function of the QCD medium (\ie, the imaginary part of the EM current-current correlation function, $\Pi_{\rm EM}$), 
$f^B (q_0;T)$ the thermal Bose distribution, evaluated in the rest frame of the medium cell moving with four-velocity,
L($M$) the lepton phase-space factor,
and $q_0$, $q$, and $M = \sqrt{q^2_0 - q^2}$ the virtual photon's energy, three-momentum and mass, respectively.
In hadronic matter, the main contribution to the EM spectral function in  the low-mass region, $M\lsim 1$ GeV, is directly related to the spectral function of the 
$\rho$ meson, ${\rm Im} \Pi_{\rm EM} = (m^{4}_{\rho}/g^{2}_{\rho})\,{\rm Im}D_{\rho}$, via the vector dominance model (much smaller contributions from the $\omega$ meson are also included).

In this paper, we utilize the EM spectral function in hot and dense hadronic matter of Refs.~\cite{Rapp:1999us,vanHees:2007th}, consisting of in-medium $\rho$- and $\omega$-meson spectral function calculated from hadronic many-body theory~\cite{Rapp:1999ej} and supplemented by a four-pion continuum with chiral mixing at masses above  $M\approx1$ GeV.
Compared to earlier studies in the SIS18 energy regime~\cite{Endres:2014zua,Galatyuk:2015pkq,Staudenmaier:2017vtq} which were based on a four-dimensional (4D) parametrization of the spectral functions, we here employ the explicitly calculated results (on a suitable grid of temperature and baryon chemical potential, as well as invariant mass and three-momentum).
The latter, in particular, provide improved accuracy at total baryon densities above saturation density, $\varrho_0$.
Since the emission rates for deconfined matter in the high-$\mu_{\rm {B}}$ region are not well known, we use the hadronic in-medium rates over the entire course of the collisions.
Due to the above mentioned quark-hadron duality of hadronic and partonic degrees of freedom in the transition region, this appears to be a fair choice.

As an alternative test case, we will also conduct calculations with the simple leading-order perturbative $q\bar{q}$ annihilation rate (neglecting any $\mu_{B}$ dependence), given by
\begin{equation}
    {\rm Im}\Pi_{\rm EM}(M) = -\frac{M^2}{12\pi} (1+\frac{\alpha_s(M)}{\pi}+...) \rm{N_c} \sum_{q=u,d}\left(e_{q}\right)^2,
\label{rate_qqbar}
\end{equation}
where N$_{\rm c}=3$ is the number of colors, and the charge sum is over the light up and down quarks.
Using this rate essentially eliminates any nontrivial medium effects in the spectral function and thus maximizes the sensitivity of the dilepton spectra to the temperature evolution of the ambient medium, encoded in the thermal Bose distribution in the emissivity.

Dilepton spectra suitable for comparison to experimental data,  as a function of invariant mass $M$, transverse momentum $p_{\rm{T}}$ (in the laboratory frame), and rapidity $y$,
$\frac{\rm{d}\it{N}_{\rm{ll}}}{\rm{d}\it{M} \rm{d}\it{p}_{\rm{T}}\rm{d}\it{y}}$, are obtained by convoluting the emissivity over the space-time evolution of the fireball.
For a hydrodynamic evolution, one can straightforwardly employ the discrete space-time cells used for its numerical evolution.
The calculation of the dilepton spectrum then reduces to a discrete sum of the dilepton rates $\frac{{\rm{d}}R_i}{\rm{d}\it{M} \rm{d}\it{p}_{\rm{T}}\rm{d}\it{y}} = 2\pi p_{\rm T} M \varepsilon_i$ from each cell $i$ at given temperature and baryon density, weighted by its four-volume $\Delta V \Delta t$ in the local restframe, 
\begin{equation}
\frac{\rm{d}\it{N}_{\rm{ll}}}{\rm{d}\it{M} \rm{d}\it{p}_{\rm{T}}\rm{d}\it{y}} = \int {\rm{d}}^4x \; \frac{{\rm{d}}R}{\rm{d}\it{M} \rm{d}\it{p}_{\rm{T}}\rm{d}\it{y}} = \sum_{i}^{\textrm{all cells}} \frac{{\rm{d}}R_i}{\rm{d}\it{M} \rm{d}\it{p}_{\rm{T}}\rm{d}\it{y}}  \Delta V \Delta t \ .
\end{equation}
To evaluate the four-momentum, $q^\mu$, of the lepton pair in the thermal restframe (figuring in the Bose distribution and the spectral function), each set of values of $M$, $p_{\rm{T}}$, and $y$ in the laboratory frame is transformed into $\vec{q}$ and $q_0$ using the cell's four-velocity $u^{\mu} = \gamma_{\textrm{cell}} \left(1, \vec{\beta}_{\textrm{cell}}\right)$, by computing the Lorentz scalar ($p\cdot u$),
\begin{equation}
q_0 = p^{\mu} u_{\mu} = \gamma_{\textrm{cell}} [ p_0 - \beta_{\rm{T, cell}} p_{\rm{T}} - \beta_{Z, \textrm{cell}} \,m_{\rm{T}} \sinh(y) ] \ .
\end{equation}
The invariant-mass spectrum is simply obtained by integrating $\frac{dN_{\rm ll}}{dM \, dp_T \, dy}$ over the pair transverse-momentum and rapidity,
\begin{equation}
\frac{{\rm d}N_{\rm{ll}}}{{\rm d}M} = \int {\rm d}p_{\rm T} \, {\rm d}y \; \frac{{\rm d}N_{\rm{ll}}}{{\rm d}M \, {\rm d}p_{\rm T} \, {\rm d}y} \ .
\end{equation}

\section{Dynamical descriptions of nuclear collisions}
\label{sec:bulk}
The standard description of the bulk dynamics of nuclear collisions at SIS18 energies to date utilizes semiclassical transport models by simulating the Boltzmann equation for the collisions of various hadrons, sometimes including the effects of nuclear mean-field interactions~\cite{Bass:1998ca,Bleicher:1999xi,Hartnack:1997ez,Hartnack:2011cn,Cassing:1999es,Ehehalt:1996uq,Teis:1996kx,Buss:2011mx}.
While these transport models can successfully describe most particle spectra and flow phenomena, the introduction of a first-order (deconfinement) phase transition is a challenging task (see, \eg, Refs.~\cite{Nara:2019qfd,Nara:2020ztb} for recent progress using an effective mean-field approach). 
In the current work we will therefore employ another approach for the dynamical description of nuclear collisions, namely ideal relativistic fluid dynamics. To illustrate the uncertainty of the role of the bulk evolution model for dilepton emission, we will compare the spectra from the hydrodynamic evolution with crossover transition to the ones from the kinetic transport approach of UrQMD. 

In this section, we briefly review the basic features of the hadronic transport model and its coarse graining (Sec.~\ref{ssec:urqmd}), the ideal-hydro evolution (Sec.~\ref{ssec:fluid}) as used in this work,  and  discuss the two EoSs corresponding to the crossover and first-order transition scenarios (Sec.~\ref{ssec:eos}). 

\subsection{UrQMD transport model}
\label{ssec:urqmd}
The UrQMD model is a microscopic transport model based on the propagation and scattering of hadrons and their resonances.
The interaction of hadrons is based on their vacuum scattering cross sections which are implemented geometrically, \ie, via a distance of closest approach criterion.
Binary interactions can either lead to elastic scattering or the excitation of either resonances or string configurations (the latter are negligible at the beam energies considered in this paper).
The corresponding reaction cross sections, as well as the different partial resonance decays widths, are taken from either experimental data~\cite{Tanabashi:2018oca} where available, or theoretical considerations (\eg, constituent quark scaling or model predictions).  

For nuclear collisions at SIS18 energies, the inclusion of nuclear mean-field interactions is considered an important part of the dynamical description of the system.
For example, the maximally achieved density is rather sensitive to these interactions, with typically $\approx30-50 \%$ smaller values in the presence of repulsive mean fields. Here we employ the standard nuclear-potential implementation of UrQMD, with a Skyrme potential where nucleons are treated as Gaussian wave packages for calculating the local net-baryon density~\cite{Hartnack:1997ez},
\begin{equation}
\centering
V_{\rm Sk}=\alpha \cdot \left( \frac{\rho _{\rm int}}{\rho _0}\right) +\beta \cdot \left( \frac{\rho _{\rm int}}{\rho _0}\right) ^{\gamma} \ .
\end{equation}
The potential parameters, $\alpha = -124$, $\beta = 71$, and $\gamma = 2$,
control the stiffness of the nuclear EoS and are chosen to best describe the different flow 
harmonics of nucleons as well as light nuclei as measured by the HADES 
experiment~\cite{Hillmann:2018nmd,Hillmann:2019wlt}.  In particular, 
the incompressibility at nuclear saturation density for the case of momentum-independent potentials as used here amounts to 380 MeV, which yields comparable results to using momentum-dependent potentials with a value of 270 MeV.

To obtain the dilepton yields from the UrQMD simulation we follow the coarse-graining approach described in detail in Ref.~\cite{Galatyuk:2015pkq}.
The space-time evolution of an ensemble of $10^4$ events at a fixed impact parameter ($b=2$ fm) is discretized into a grid of individual cells of size $\Delta V_4 = 1 {\rm fm}^3 1 {\rm fm}/c$.
For each time step the UrQMD particles are smeared with a three-dimensional Gaussian around their position and filled into the cells.
As thermalization criterion we require the majority ($>70\%$) of nucleons in a given cell to previously have experienced at least three interactions with other particles (elastic or inelastic, including resonance decays).
Cells that do not fulfill this criterion are not included in the calculation of dilepton emission.
This criterion removes most cells from the early time steps  while the two nuclei still penetrate each other, roughly for the first $\approx 7$ fm/$c$ after the nuclei start to touch. This procedure is corroborated by recent work~\cite{Wiest:2021} which shows that after this time the energy-momentum tensor is indeed near isotropic using the criterion developed in Ref.~\cite{Florkowski:2010cf}.
The baryon density inside each cell is determined by evaluating the baryon four-current, $j^{\mu}$; the latter also yields the cell's velocity needed to 
boost the hadron (and dilepton) momenta into the local rest-frame of the cell.
The cell temperature is then obtained from the slope of an exponential fit to the transverse-momentum spectrum of pions inside the cell.

Temperature, cell velocity and baryon density serve as input parameters to the local dilepton rate.
The spectrum of thermal dileptons is determined by summing over the contribution of all cells (weighted by the cell four-volume). 
To compare the coarse-grained UrQMD result on an equal footing to the space-time evolution from the fluid dynamics simulations, cells with temperatures below 50 \mev\ or an energy density below $0.9 \epsilon_0$ are not considered for the calculation of dilepton spectra.

\subsection{Ideal fluid dynamics}
\label{ssec:fluid}
Fluid dynamics has been used early on to simulate low-energy nuclear collisions~\cite{Scheid:1974zz,Hofmann:1976dy,Stocker:1981zz} and has by now become the state-of-the-art model in the ultra relativistic regime at the LHC and RHIC and down to SPS beam energies.
A main advantage of fluid dynamics is that it is directly based on the EoS of the medium, and thus its manifestation in observables can be readily studied. 
In the present context, the inherent assumption of (near) local-equilibrium of the medium
cells allows for a straightforward calculation of thermal dilepton spectra as discussed in the previous section.

The equations of ideal fluid dynamics, under the conservation of the baryon number four current, are given by
\begin{equation}
\partial_{\mu} T^{\mu \nu} =0\ ,
\end{equation}
and local flavor conservation,
\begin{equation}\label{consr}
\partial_{\mu} j^{\mu}=0 \  , 
\end{equation}
essentially expressing the conservation of energy momentum and net baryon number.
In principle, the fluid's energy-momentum tensor can be extended to higher-order corrections due to local momentum anisotropies which would lead to the formulation of an effective viscous fluid dynamics model.
Since in the current work we are mainly interested in the bulk evolution of the system driven by low-momentum particles, a nonviscous evolution is usually a fair approximation.
This also saves us the introduction of additional transport parameters which, at high $\mu_{\rm B}$, have not been reliably assessed to date.
For example, in Ref.~\cite{Karpenko:2015xea} empirical hydrodynamics-based extractions of $\eta/s$ from Au-Au collisions in the RHIC energy regime ($\sqrt{s}=7.7-200$ GeV) have found a moderate increase toward lower energies, from about 0.1 to 0.2.
However, such extractions are beset with significant uncertainties in other ingredients of the simulations, such as the initial-state configurations, formation times, or the temperature dependence of $\eta/s$.
In a transport-based hadron resonance matter study~\cite{Rose:2017bjz} it was found that $\eta/s$ is rather constant with chemical potential and even tends to decrease slightly with $\mu_{\rm B}$ for temperatures pertinent to the present study, $T \lsim 80$ MeV.
On the other hand, perturbative studies in deconfined matter usually predict higher $\eta/s$ values when the quark fraction is increased due to reduced color-Casimir factors in quark- relative to gluon-induced scatterings.
Clearly, more careful studies of viscosity effects in hydrodynamic simulations of heavy-ion collisions at relatively low energies are needed and how they impact the emission characteristics of EM radiation.
This is, however, beyond the scope of the present paper.

The fluid dynamic equations are solved on a 3+1-dimensional Euclidean grid with cubic cells of size $\Delta x=0.2$ fm and with a time step of $\Delta t= 0.08$ fm/$c$, to adequately ensure current conservation, using the well-tested SHASTA code~\cite{Rischke:1995ir}. 
In particular, it has been extended with a local finite-range term which is mandatory for the treatment of the spinodal region in the first-order scenario~\cite{Steinheimer:2012gc}.
The initialization of the fluid dynamical simulation is done with two Lorentz-contracted, properly normalized, Wood-Saxon distributions in the center-of-mass frame of the collision, just prior to contact.
As fluid dynamics assume instant local equilibrium in every cell, the energy, momentum, and baryon number of the two incoming nuclei is very rapidly stopped in the simulation.
For higher beam energies this can lead to issues of overestimated stopping.
However, it is known that the amount of entropy per baryon produced due to the shock heating is comparable to transport simulations~\cite{Arsene:2006vf}.
Transport simulations also suggest that the assumption of thermalization is justified after a few fm/$c$~\cite{Steinheimer:2016vzu}. 

In the present paper we will focus on the effect of a first-order phase 
transition on the bulk evolution to estimate the emerging dilepton production; for that purpose we restrict ourselves to the evolution of event-averaged densities and temperatures, as was also done in our previous studies using the coarse-grained transport approach. The implementation of density fluctuations would involve additional model uncertainties while likely not improving the accuracy of our calculations in this exploratory study.

\subsection{Equation of state}
\label{ssec:eos}
The EoS relates the pressure of the fluid, $P \equiv P(\epsilon, {\rm n_B})$ to other thermodynamic quantities, \ie, energy and net-baryon density. 
Since first-principle lQCD computations for the EoS at large $\mu_{\rm B}$ are not available, we resort to a model calculation constrained by phenomenology in terms of the chiral mean-field (CMF) model as developed by the Frankfurt group~\cite{Steinheimer:2011ea}, which in essence is a chirally symmetric linear $\sigma$-$\omega$ model with an explicit mass term for the baryons which preserves chiral symmetry.
This choice is based on the idea of the parity doublet model of ground-state baryons and their excited opposite-parity partner states~\cite{Detar:1988kn,Hatsuda:1988mv,Zschiesche:2006zj,Sasaki:2010bp}.
In this scenario, chiral symmetry restoration is realized through a degeneracy where the mass of the excited state drops to the one of the ground state.
This mechanism is supported by baryon correlation functions computed in QCD~\cite{Aarts:2017rrl}, as well as by dilepton spectra in ultrarelativistic HICs where the $\rho$ width strongly increases but its in-medium mass is essentially stable and degenerates with a dropping $a_1$ mass~\cite{Hohler:2013eba}. 
The CMF model also includes a smooth transition to a deconfined quark phase, where the  thermal contribution of the quarks is added to the grand canonical potential in analogy to Polyakov-loop Nambu-Jona-Lasinio models~\cite{Fukushima:2003fw,Ratti:2005jh}, while hadronic degrees of freedom are gradually suppressed through an excluded-volume prescription effectively acting on the chemical potential of all hadrons~\cite{Steinheimer:2010ib}.
In Ref.~\cite{Mukherjee:2016nhb} the parameters of the quark phase were chosen to reproduce a smooth crossover from hadronic to quark degrees of freedom, consistent with lQCD findings at $\mu_{\rm B}=0$~\cite{Borsanyi:2013bia}.
As a result, the chiral transition remains a crossover down to temperatures of $T>50$ \mev, below which a weak first-order phase transition develops.
It was also shown~\cite{Mukherjee:2017jzi} that this EoS is consistent with constraints from neutron star mass-radius relations and neutron star mergers; its incompressibility at nuclear saturation density is compatible with that of the UrQMD model with mean fields as described above.
\begin{table}[tbh]
\center
\begin{tabular}{|c|c|c|}
\hline
Parameter & Crossover & First-order Phase Transition \\ \hline \hline
$g_{q \sigma}$& 2.5 & 5.2 \\ \hline
$g_{s \zeta}$& 2.5 & 4.2 \\ \hline
$g_{q \omega}$& 0.0 & 3.0 \\ \hline
\end{tabular}
\caption{Variations of parameters in the Frankfurt CMF model that allow to change from crossover (for $T > 50$ MeV) to a first-order phase transition at higher $T$.}
\label{tab:param}
\end{table}
\begin{figure}[tbh]
\centering
   \includegraphics[width=0.7\textwidth]{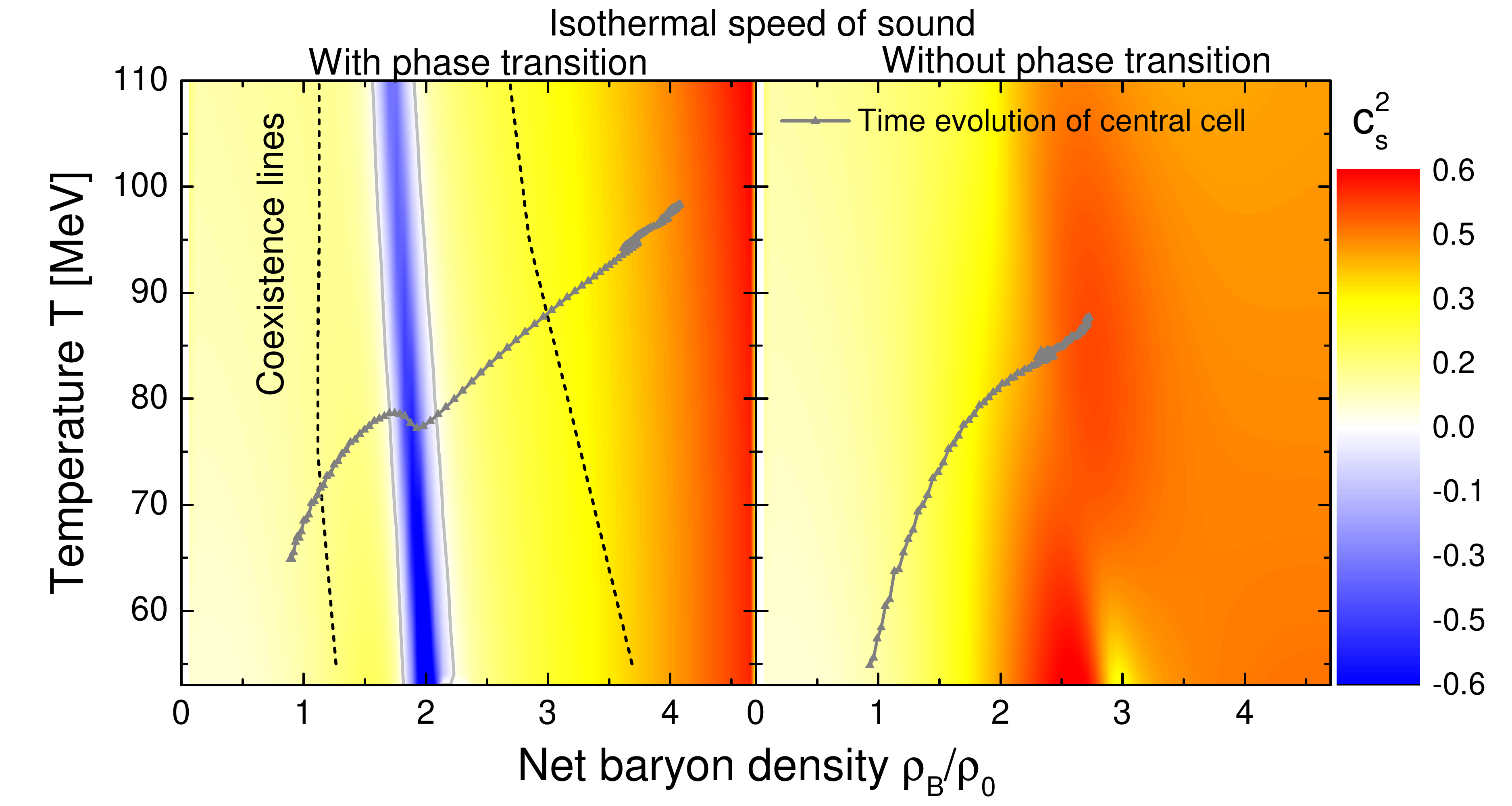}
   \caption{
   (Color online) Speed of sound for first-order EoS (left panel) and the crossover EoS (right panel) in the Frankfurt CMF model. For the first-order scenario the mechanically unstable region is shown in blue color, encompassed by the spinodal lines (dashed lines). The coexistence region, where metastable solutions of phases can exist, is indicated by the dashed lines. To illustrate the approximate dynamical evolution path in a HIC, the time evolution of the central cell (after $t = 7$ fm/$c$) in the full fluid dynamical evolution is shown as a gray line with symbols in each scenario.
   }
\label{fig:cs2a}
\end{figure}

Since, in the current work, we want to investigate the effects of a first-order phase transition in nuclear collisions, the parameters of the model can be modified for this transition to develop at higher temperatures, although the pertinent change in parameters will generally create a tension with lQCD EoS near vanishing $\mu_{\rm B}$.
However, the aim of this paper is to find experimental observables which are sensitive to such a phase transition which is complementary to the theoretical analyses.
The changes in parameters from the crossover to the first-order transition are summarized in Table~\ref{tab:param}; they amount to changing the coupling strengths of the free quarks to the scalar fields (nonstrange $\sigma$ and strange $\zeta$) as well as the repulsive vector field $\omega$ (cf.~Ref.~\cite{Mukherjee:2016nhb} for further details).
We note that these couplings do not change the nuclear matter properties of the model but do affect the phase structure of the matter at high density.

\begin{figure}[tbh]
\centering
   \includegraphics[width=0.5\textwidth]{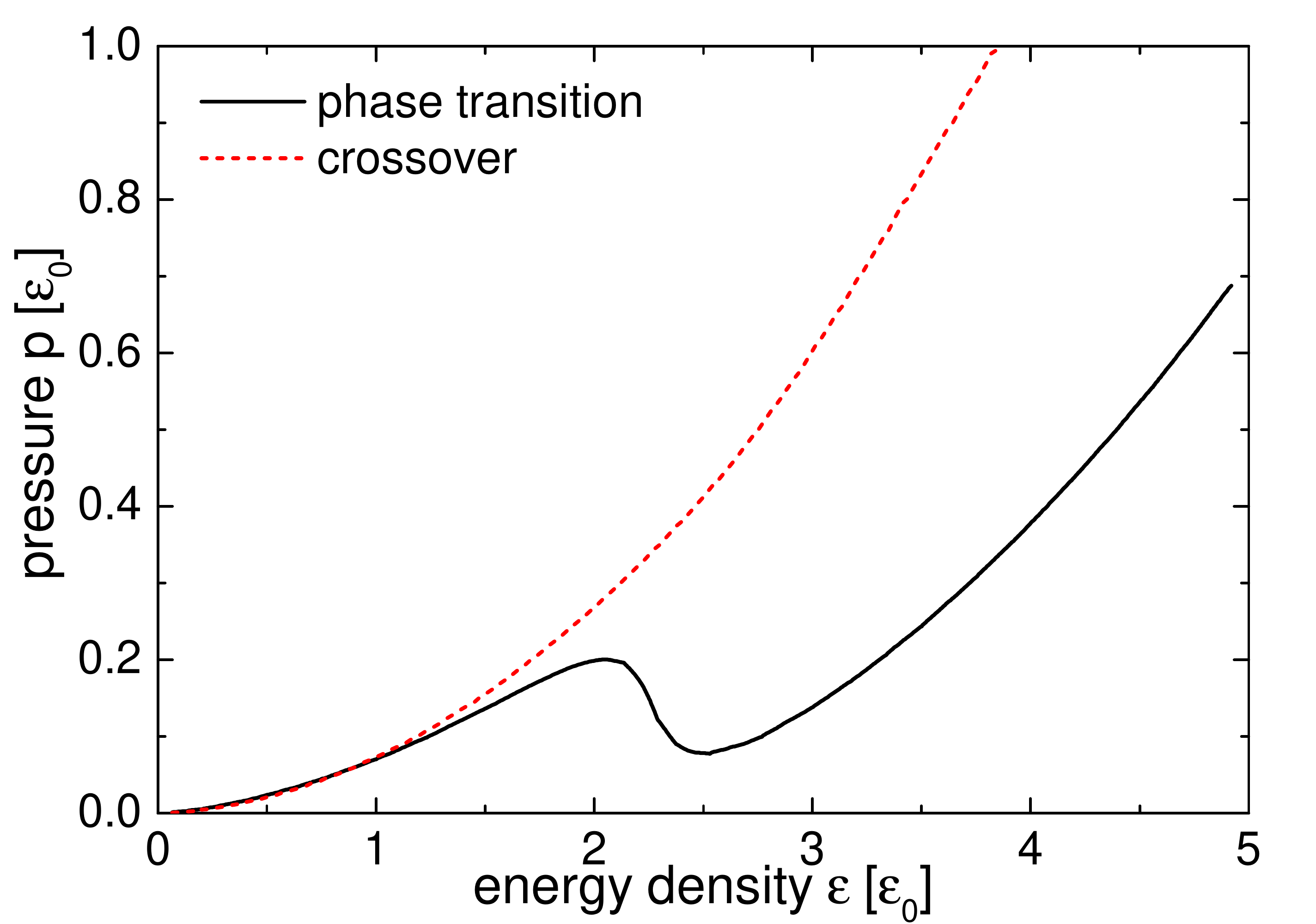}
   \caption{
   (Color online) Pressure as a function of energy density along a line of constant entropy per baryon, $S$/A.
   }
\label{fig:cs2b}
\end{figure}

To illustrate the properties of the two different EoSs the isothermal speed of sound for both parametrizations is shown in Fig.~\ref{fig:cs2a}.
With our parameter choice a smooth crossover, relatively stiff EoS (Fig.~\ref{fig:cs2b}) can be changed to a first-order phase transition in the density range relevant for heavy-ion collisions at SIS18 energies.
The width of the transition is illustrated in the left panel of Fig.~\ref{fig:cs2a} by the mechanically unstable phase of the transition with negative $c_{\rm{s}}^{2}<0$ (as blue band) and by the spinodal lines (dashed black lines) which mark the appearance of metastable phases in the phase diagram. The resulting latent heat amounts to a discontinuity of about 85\% in the entropy density (relative to the mean value in the transition). 

The approximate dynamical evolution of the system is indicated in the left and right panel of Fig.~\ref{fig:cs2a} by the trajectories of the central cells of the fluid dynamical simulations for central ($b=2$\,fm) Au-Au collisions at a beam energy of $E_{\mathrm{lab}}=1.23$~\agev.
These trajectories essentially correspond to an isentropic expansion after the largest compression has been reached.
The full space-time evolution will be discussed in the following sections.

\section{Fireball Evolution and Dilepton Spectra in 1.23 $A$GeV Au-Au Collisions}
\label{sec:results}
In the following we first compare the results from the different dynamical simulations of \auau\ collisions at $E_{\mathrm{lab}}=1.23$ \agev, specifically the final-state pion spectra (Sec.~\ref{ssec:pions}) and the evolution of temperature and baryon density (Sec.~\ref{ssec:Tmu-evo}), and then analyze the resulting 
dilepton production (Sec.~\ref{ssec:dl-spec}). 
%
\subsection{Comparison of pion spectra}
\label{ssec:pions}
%
In order to understand the role of EM probes (dilepton emission) of the matter in nuclear collisions, one first needs to achieve some control over the role of hadronic observables.
Most measurements in high-energy HICs relate to single-hadron yields and spectra as well as pertinent correlation functions.
The most widely used species are pions which are abundantly produced and reasonably simple to measure.
Since the EoS affects the created volume and lifetime of the system, one may ask to what extent the final-state hadron/pion spectra are sensitive to it.
Since the hadrons are emitted, by definition, at the point of decoupling, the measured spectra provide a snapshot at thermal freeze-out (where elastic interactions cease), while the hadron yields are usually determined earlier, at chemical freeze-out (where inelastic particle-number changing interactions cease). 
At this time the system has usually reached the stage of a hadronic medium which is common to most dynamical models, including the fluid dynamical and transport approaches used in the present work.
Even angular correlations do not necessarily have a direct connection to the EoS~\cite{Steinheimer:2014pfa} at SIS18 energies, possibly requiring more sophisticated analysis methods~\cite{Pang:2016vdc,Du:2019civ}.  

\begin{figure}[tb]
\centering
   \includegraphics[width=0.5\textwidth]{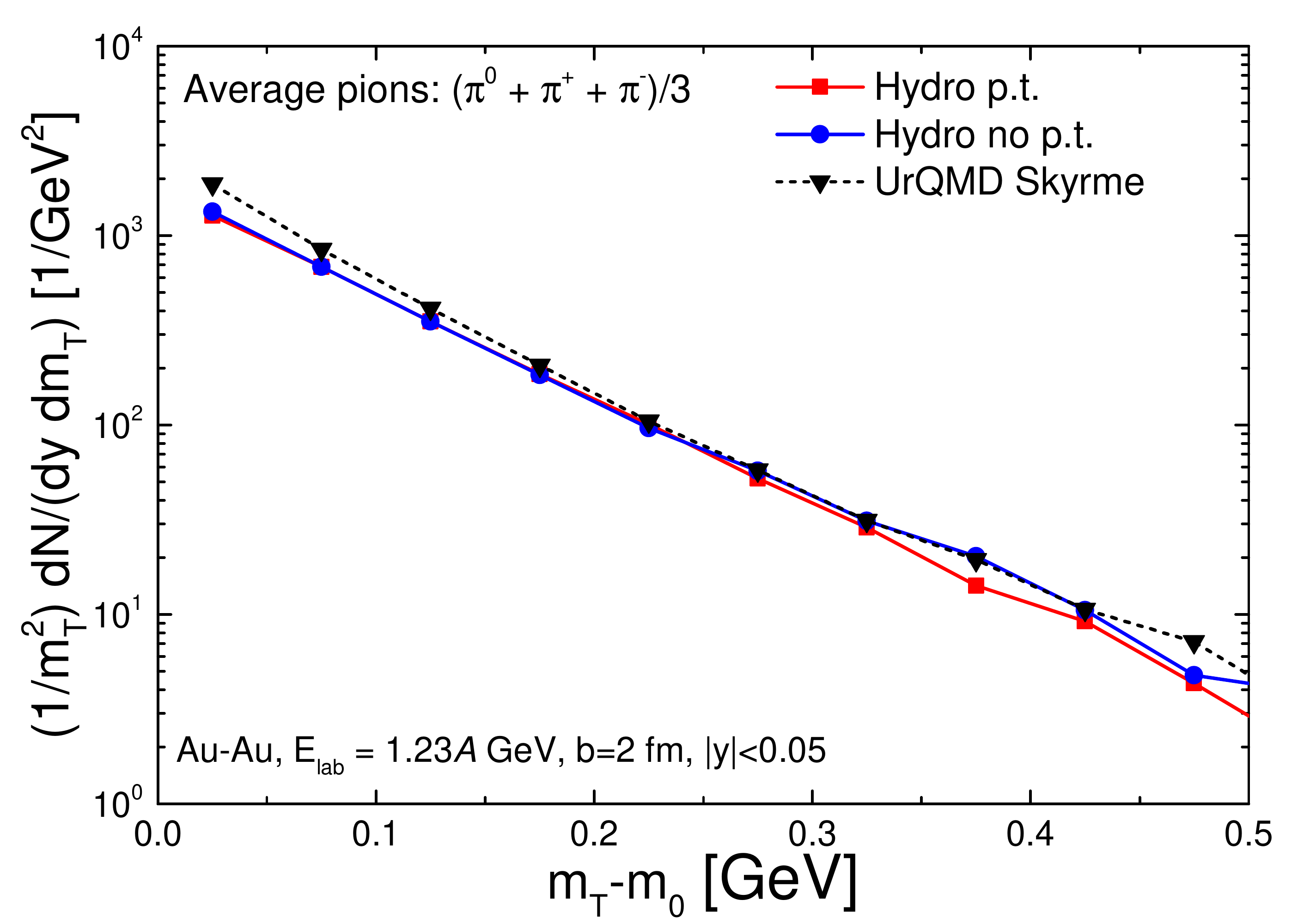}
   \caption{(Color online) Comparison of the final pion transverse-mass
    spectrum from the three bulk evolution models used in this work for $b=2$ fm 1.23 $A$GeV Au-Au collision at  midrapidity ($|y|<0.05$). The two fluid dynamical results are almost identical while the transport model shows a small deviation from those at small $m_{\rm T}$.}
\label{fig:pions}
\end{figure}

At the minimum we want to ascertain that the bulk evolution models employed here give a reasonable description of pion production.
Thus, we have extracted the pion transverse-mass spectra around midrapidity from the UrQMD-Skyrme mode as well as from the fluid-dynamical simulations, cf.~Fig.~\ref{fig:pions}.
For the fluid dynamics we have employed the standard Cooper-Frye method~\cite{Cooper:1974mv} to compute the spectra on an isoenergy density hypersurface at $e=3 \cdot e_0$ (where $e_0=0.15$ GeV/fm$^3$ is the nuclear ground-state energy density), with the same resonance content as in the UrQMD model.
This density defines our chemical freeze-out condition (which is not equal to the thermal freeze-out condition).
Its choice is based on properly describing the measured pion yields.
All three model variants give approximately the same yield and shape of the pion spectra, which provides a reasonable starting point for a meaningful comparison of dilepton emission. 
The small difference between the fluid-dynamical and the transport simulations, with slightly harder spectra for the former, is presumably related to a stronger collectivity generated by hydrodynamics in the early phases.
One should also note that a recent comparison of HADES data found that the UrQMD model gives a reasonable description of the experimental results~\cite{Adamczewski-Musch:2020vrg}. 

As the subsequent evolution from chemical to thermal freeze-out continues to produce appreciable dilepton radiation, we should ensure that the number of pions (as the main catalysts of dilepton emission) is approximately conserved during the later part of the expansion.
It turns out that this is the case for the first-order scenario, but for the crossover the total 
pion number drops from $\approx 11$ to $\approx 7$.
We correct for this following the standard procedure employed in high-energy HICs by introducing a pion chemical potential that increases approximately linearly with decreasing $T$ from chemical to thermal freeze-out~\cite{Rapp:2002fc,Teaney:2002aj,Hirano:2002ds}.
For the case at hand, we find $\mu_\pi^{\rm fo}\simeq 22$ MeV at $T_{\rm fo}=56$ MeV for the crossover scenario at thermal freeze-out, so that the pertinent fugacity factor, $z_\pi = \exp(\mu_\pi/T)$ amounts to $\approx 11/7$.
Note that the fugacity factor also applies to hadronic resonances with strong decay branchings into pions according to the law of mass action, \eg, $\mu_\Delta = \mu_{\rm N} + \mu_\pi$ corresponding to $\Delta \leftrightarrow \rm {N} + \pi$ with $\approx 100\%$ branching ratio.
The additional fugacity also figures in the EoS.
However, in the nondegenerate limit, both pressure and energy density of the respective particle components are affected equally, and thus their effect on the EoS (or speed of sound) approximately cancels.
Consequently, the hydrodynamic evolution is affected very little~\cite{Teaney:2002aj} by the fugacities, and we can therefore safely neglect this effect here. 
For the UrQMD evolution we adopt the pion chemical potentials as found in our previous work~\cite{Galatyuk:2015pkq}.
In that work we have also determined that, based on the different dilepton production channels (NN Bremsstrahlung, resonance Dalitz decays and $\pi\pi$ annihilation or $\rho$ decays), the average number of pions involved is $\kappa=1.12$.
Thus, the local dilepton rates in the UrQMD and hydro-crossover scenarios are augmented by an effective dilepton fugacity of $z_\pi^\kappa$.

%

\subsection{Temperature and density evolution}
\label{ssec:Tmu-evo}
\begin{figure}[tb]
\centering
   \includegraphics[width=0.3\textwidth]{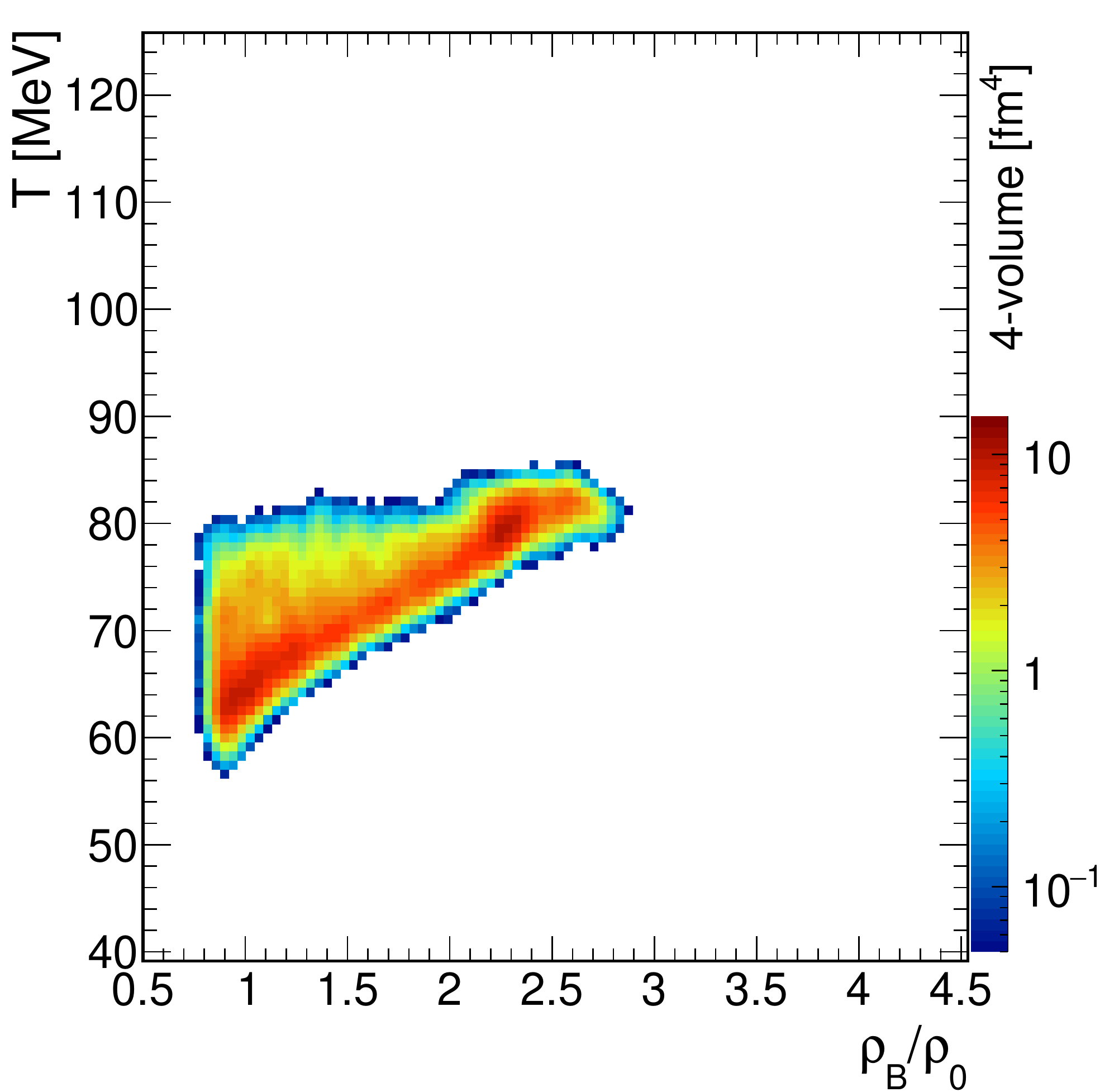}
   \hspace{0.5cm}
   \includegraphics[width=0.3\textwidth]{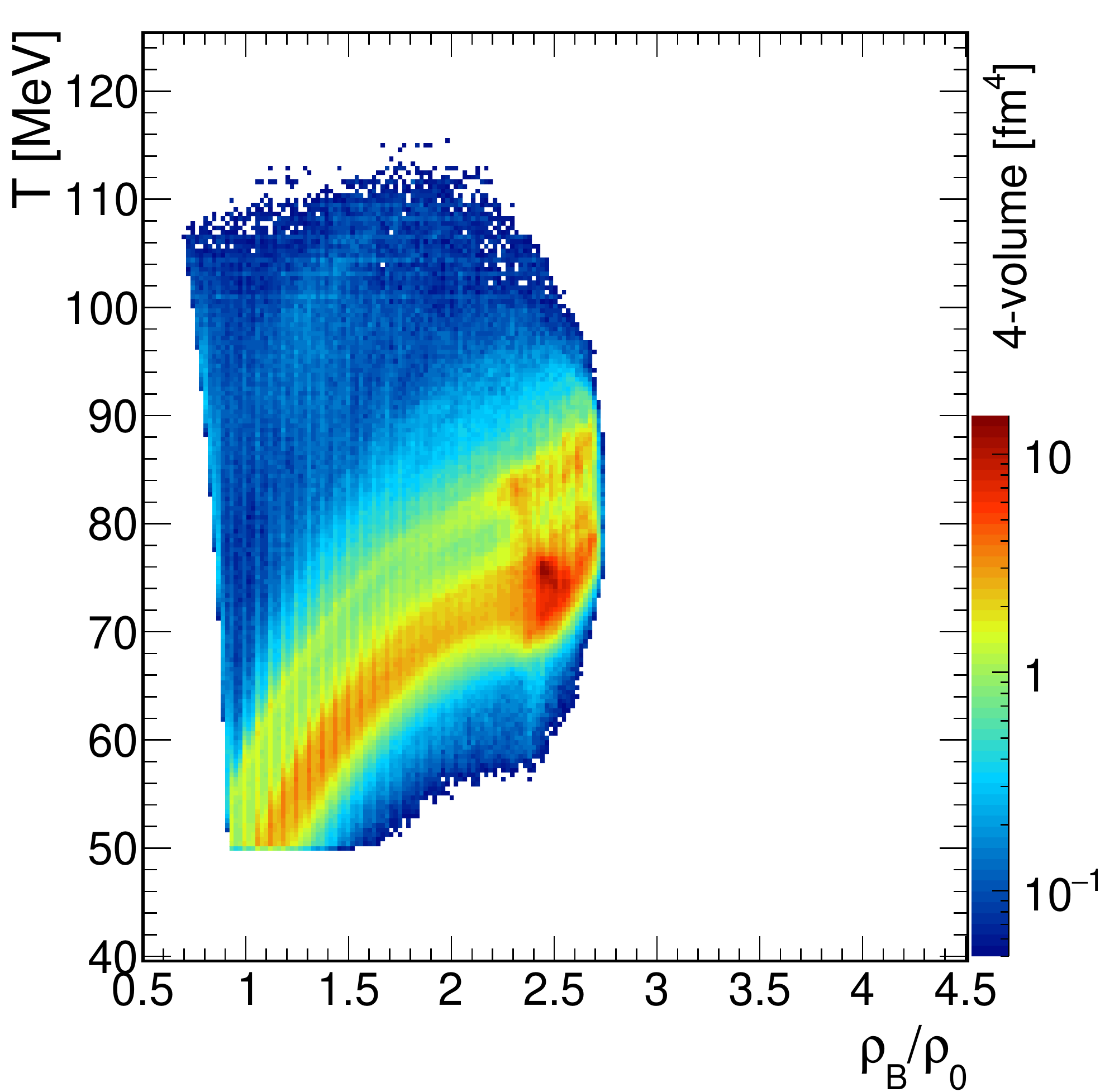}
   \hspace{0.5cm}
   \includegraphics[width=0.3\textwidth]{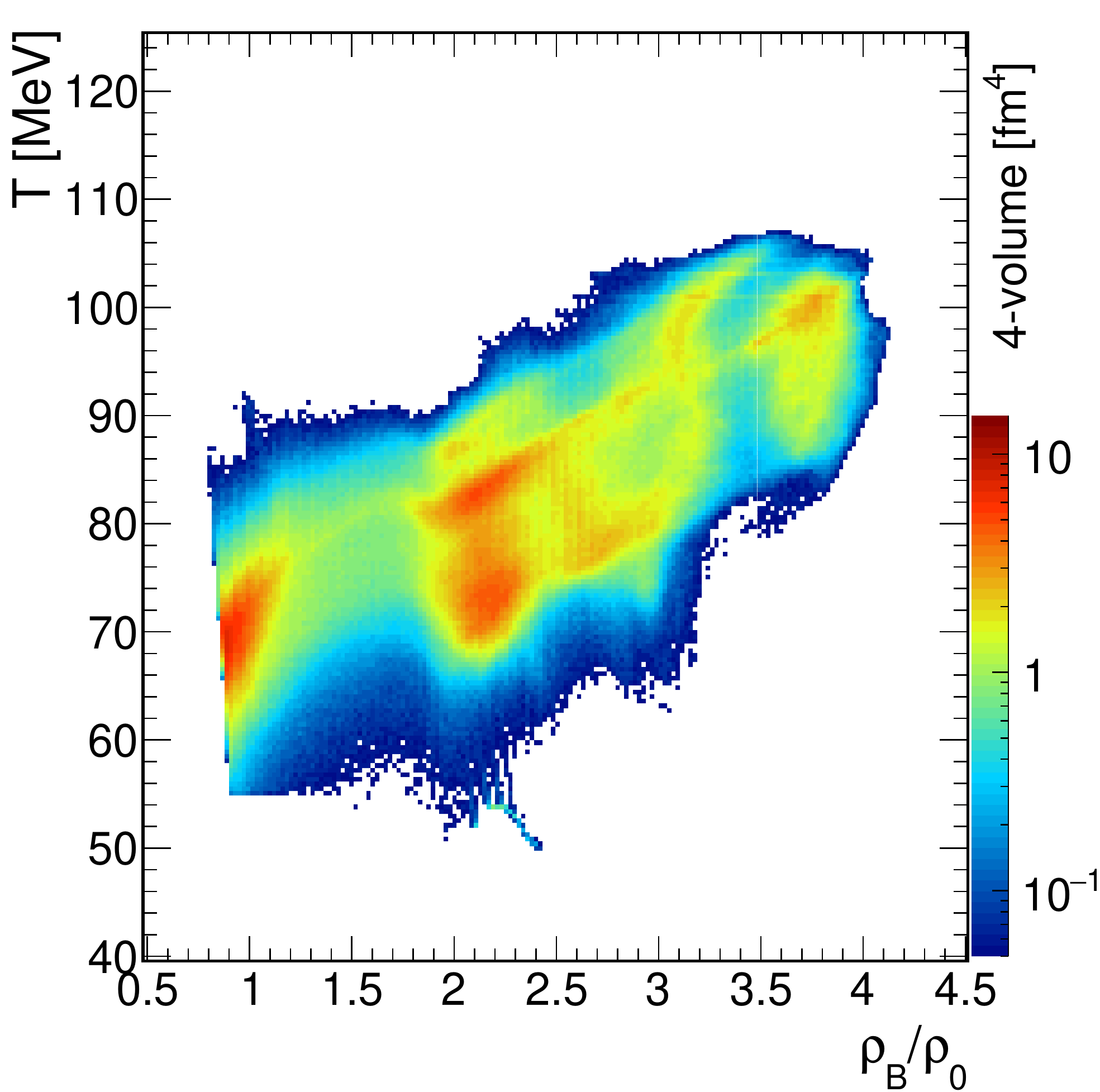}
   \caption{
   (Color online) Distributions of the four-volume weights in fireball temperature and baryon density in central \auau\ collisions at 1.23 \agev\ extracted from the coarse-grained UrQMD transport model (left panel) and ideal-hydro evolutions with crossover EoS (middle panel) and first-order EoS (right panel). Contributions from the first 7 fm/$c$ after initial impact are not included.
   }
\label{fig:3dhisto}
\end{figure}

After benchmarking the final-state pion spectra, the next step is to inspect the bulk evolution in terms of the extracted temperature ($T$) and baryon density ($\varrho_{\rm B}$) profiles, which are the key ingredients to the calculation of dilepton spectra in the following section. 
Toward this end we fill 3D histograms in the $T$-$\varrho_{\rm B}$ plane representing the space-time weight of each value of the thermodynamic variables.
For the fluid dynamical evolutions this is directly obtained from the discrete cells used in the numerical simulations, while for the coarse-grained transport we employ the method laid out in Sec.~\ref{ssec:urqmd}; the results are collected in the three panels of Fig.~\ref{fig:3dhisto}.
For this figure we have excluded any contributions from times before 7 fm/$c$, since it is not clear whether the matter at such early times can be considered thermalized, as alluded to in Sec.~\ref{ssec:urqmd} (the coarse graining only yields about 10\% of the total amount of radiated lepton pairs from this period).
We will return to a more quantitative assessment of the contributions of these cells to dilepton spectra from the hydro evolutions in Sec.~\ref{ssec:highT} below. 

For most of the four-volume the coarse-grained transport and  fluid dynamical descriptions without a phase transition give comparable results: The majority of the system reaches temperatures and baryon densities of up to 85 MeV and $3 \varrho_{0}$, respectively, during the hot and dense stage of the collisions.
However, the spread in temperature is significantly higher for the crossover hydro evolution, especially on the high end with temperatures reaching near 110 MeV (albeit with a small space time weight).
For the hydro evolution with a first-order phase transition in the EoS, substantially higher densities of up to $4 \varrho_0$ are reached, and also cells with a temperature of up to 105 MeV have an appreciable space-time weight, which is much larger than in the crossover scenario.
This is readily understood as the heating in the fluid simulation occurs as a result of the created shock, as the two nuclei hit head on.
As the matter enters the phase transition region, it becomes much more compressible due to the drop in the speed of sound (recall Fig.~\ref{fig:cs2a}), resulting in larger initial temperatures and densities. Furthermore, during the evolution of the system, the fireballs of the UrQMD model and crossover fluid model exhibit a rather smooth decrease of temperature and density, while the evolution with phase transition develops separated areas in the $T$-$\varrho_{\rm B}$ plane corresponding to the two coexisting phases.
The thermal phase-space region in between these two phases is ``deserted" as the matter here is mechanically unstable and exponentially decays into any of the two coexisting phases.

\begin{figure}[tb]
\centering
   \includegraphics[width=0.325\textwidth]{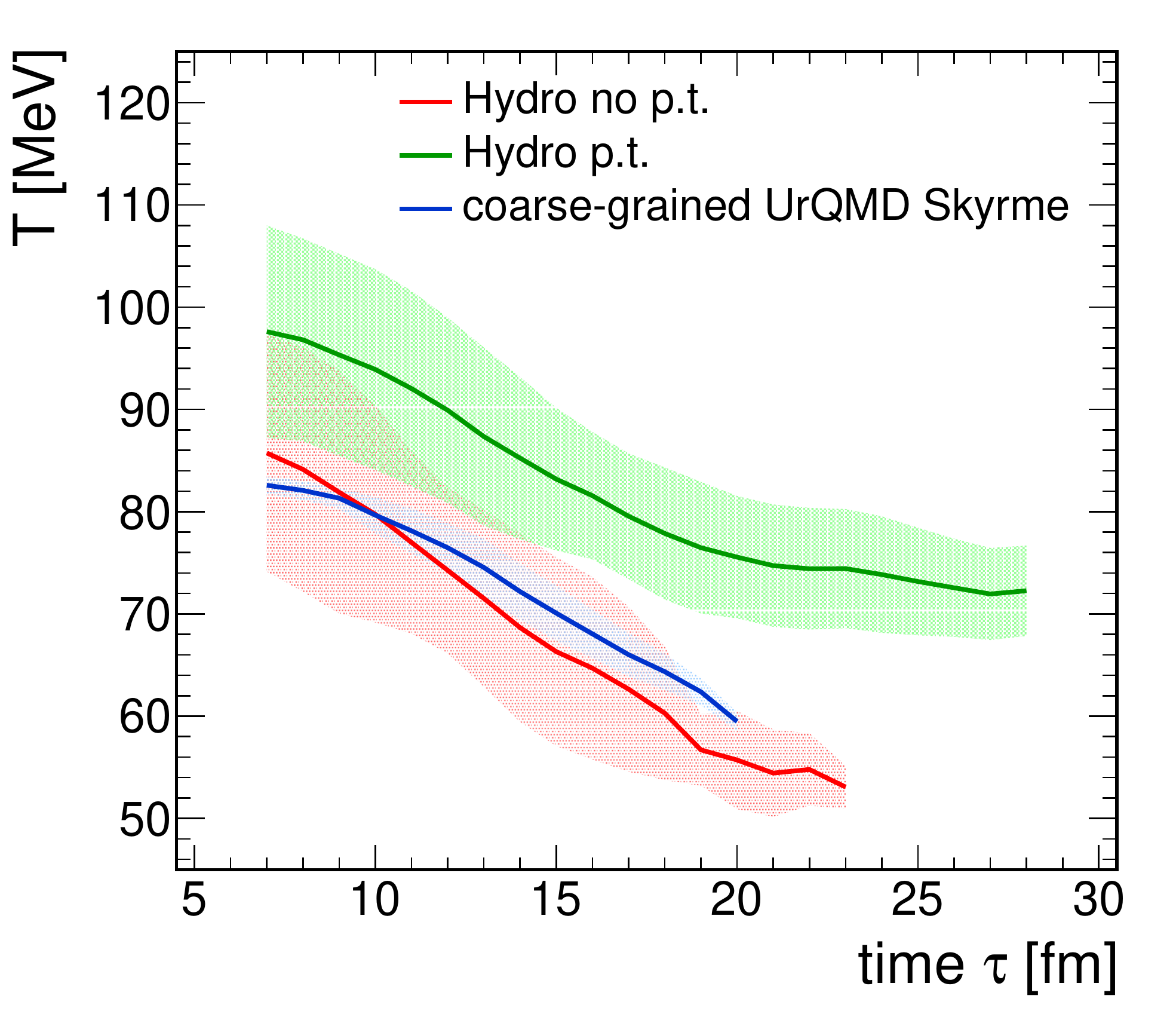}
   \includegraphics[width=0.325\textwidth]{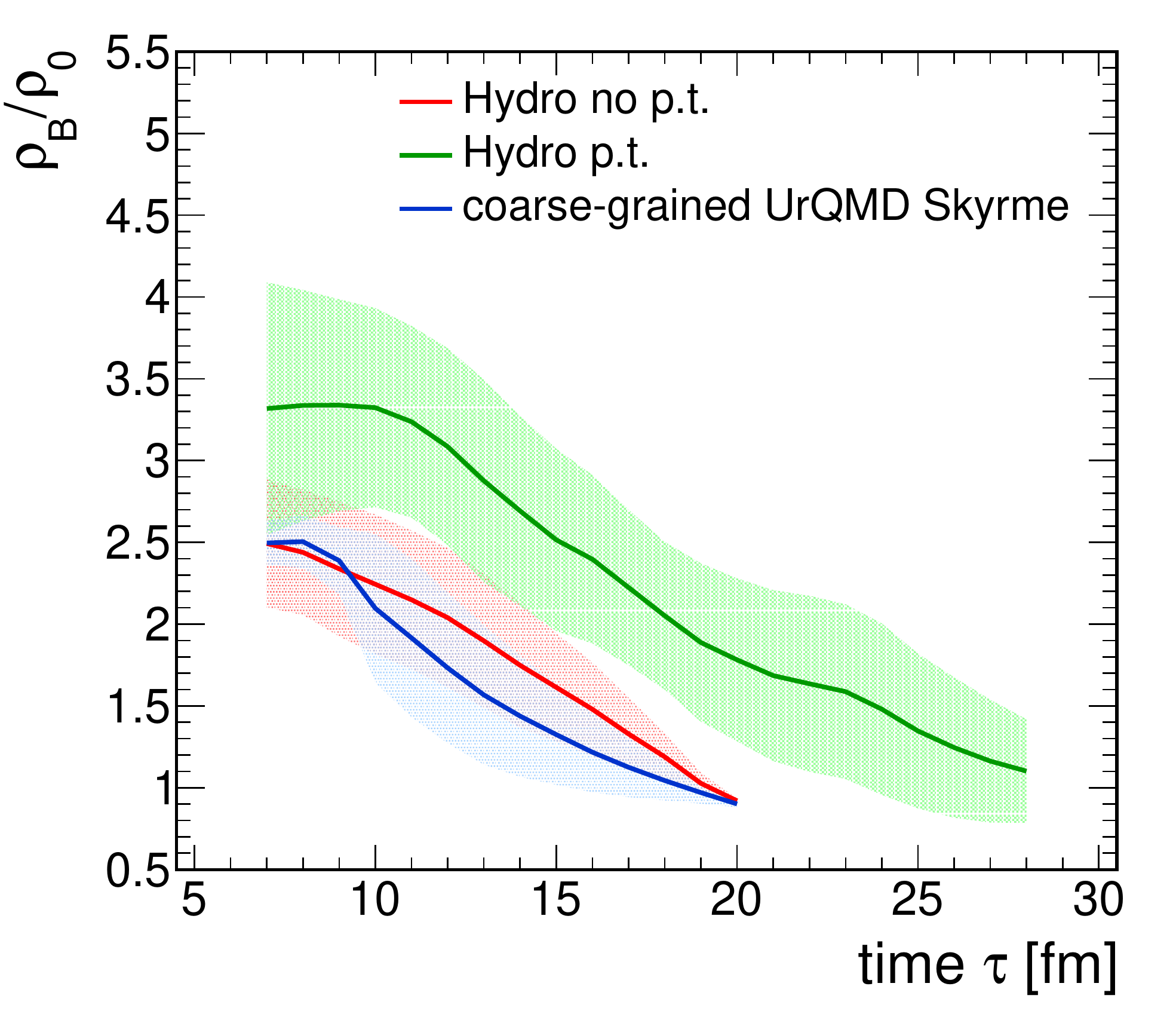}
   \includegraphics[width=0.325\textwidth]{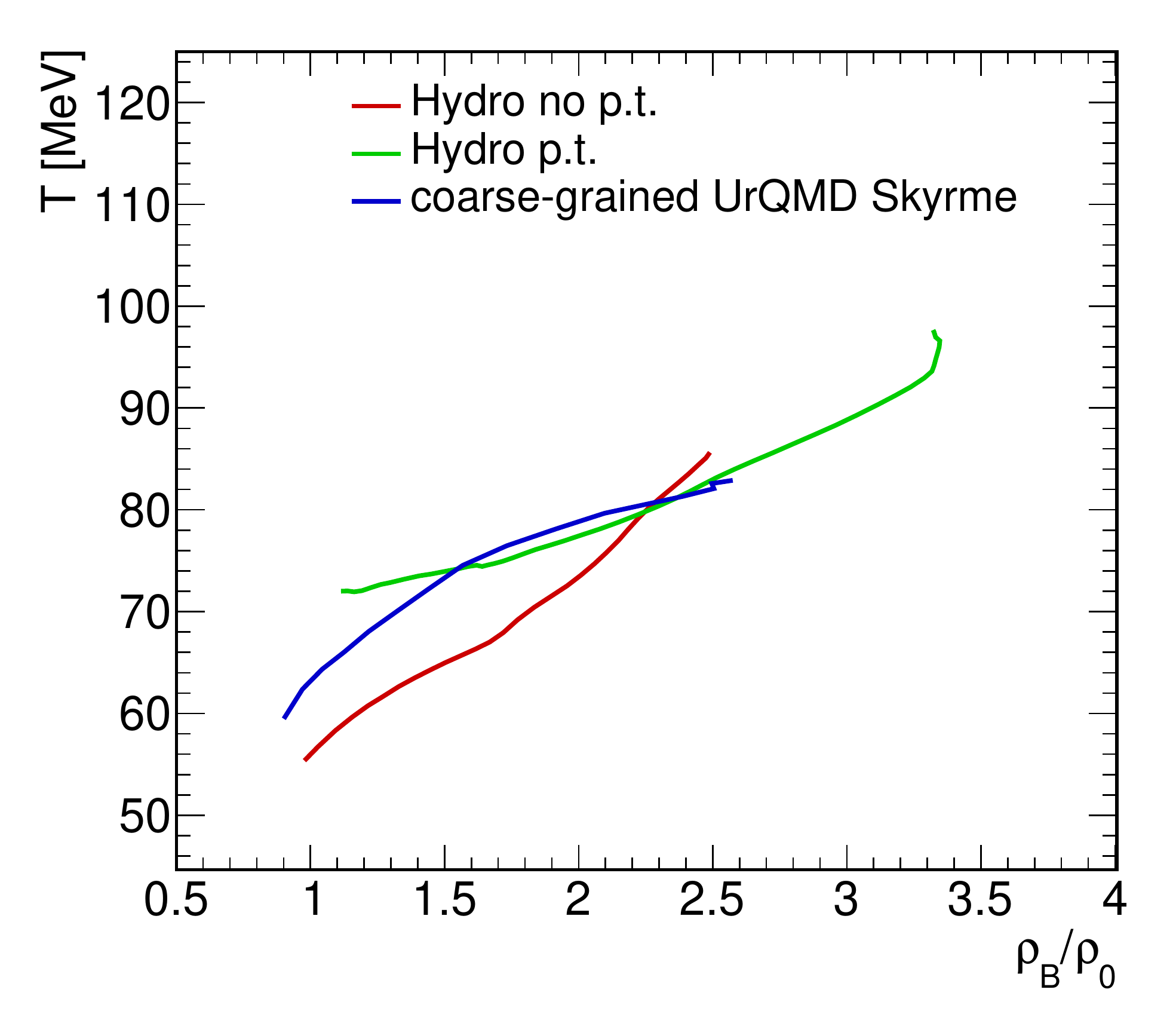}
   \caption{
   (Color online) Time evolution of the fireball temperature (left panel) and density (middle panel) in \auau\ collisions at 1.23 \agev\ extracted from the coarse-grained UrQMD transport model simulation (blue curve), ideal hydrodynamics with either crossover (red curve) or first-order EoS (green curve). The bands correspond to the second moment of the corresponding distributions.
   The right panel shows fireball trajectories in the temperature–density plane, defined by the most probable value of $T$ and $\varrho_{\rm B}$ at a given time, as extracted from the three simulation models (same color code as in the other panels).
   }
\label{fig:temp_dens_1d_7fm}
\end{figure}

For a more inclusive representation of our results, we plot the time evolution of the mean temperature and baryon density in Fig.~\ref{fig:temp_dens_1d_7fm}, while quantifying of the spread in terms of a one-sigma standard deviation indicated by the colored bands around the mean-value curves.
As expected from the four-volume histograms, the coarse-grained transport and the crossover fluid evolution of the collision show rather good agreement, both in terms of the average as well as the system lifetime which terminates after about 20 fm/$c$ (when all cells have cooled below 50 \mev\ in temperature and $0.9 \epsilon_0$ in energy density); the spread is somewhat smaller for the transport model, some of which may be caused by the coarse-graining procedure to begin with.
For the first-order hydro evolution, the temperature and baryon density are significantly higher (again in line with the 4D histogram),  with slightly larger spread than for the crossover EoS, and a markedly longer lifetime by almost 10~fm/$c$.  
In the early stages, the spread in temperatures is, in fact, slightly larger in the crossover fluid simulation compared to the phase transition case, as the very hottest cells of all simulations are created in the former (recall Fig.~\ref{fig:3dhisto}).
The origin and consequences of these hot cells will be discussed in Sec.~\ref{ssec:highT}.
Since it is only a small fraction of cells (less than 1\%), the mean temperature value is not significantly affected by them. However, dileptons at larger invariant masses ($M > 1.2$ \gev) become increasingly sensitive to the hottest temperatures reached during the collision due to the thermal factor in Eq.~(\ref{emisthermeq}).

Finally, in the right panel of Fig.~\ref{fig:temp_dens_1d_7fm} we compare the  trajectories of the mean temperature and density for the three different models in the $T$-$\varrho_{\rm B}$ plane.
The differences are likely a consequence of the different effective degrees of freedom in the three models.
Despite these differences the trajectories give a good indication of the area in the phase diagram that is probed by the central collisions at the SIS18 accelerator.

%

\subsection{Dilepton Spectra}
\label{ssec:dl-spec}
\begin{figure}[tb]
\centering
   \includegraphics[width=0.42\textwidth]{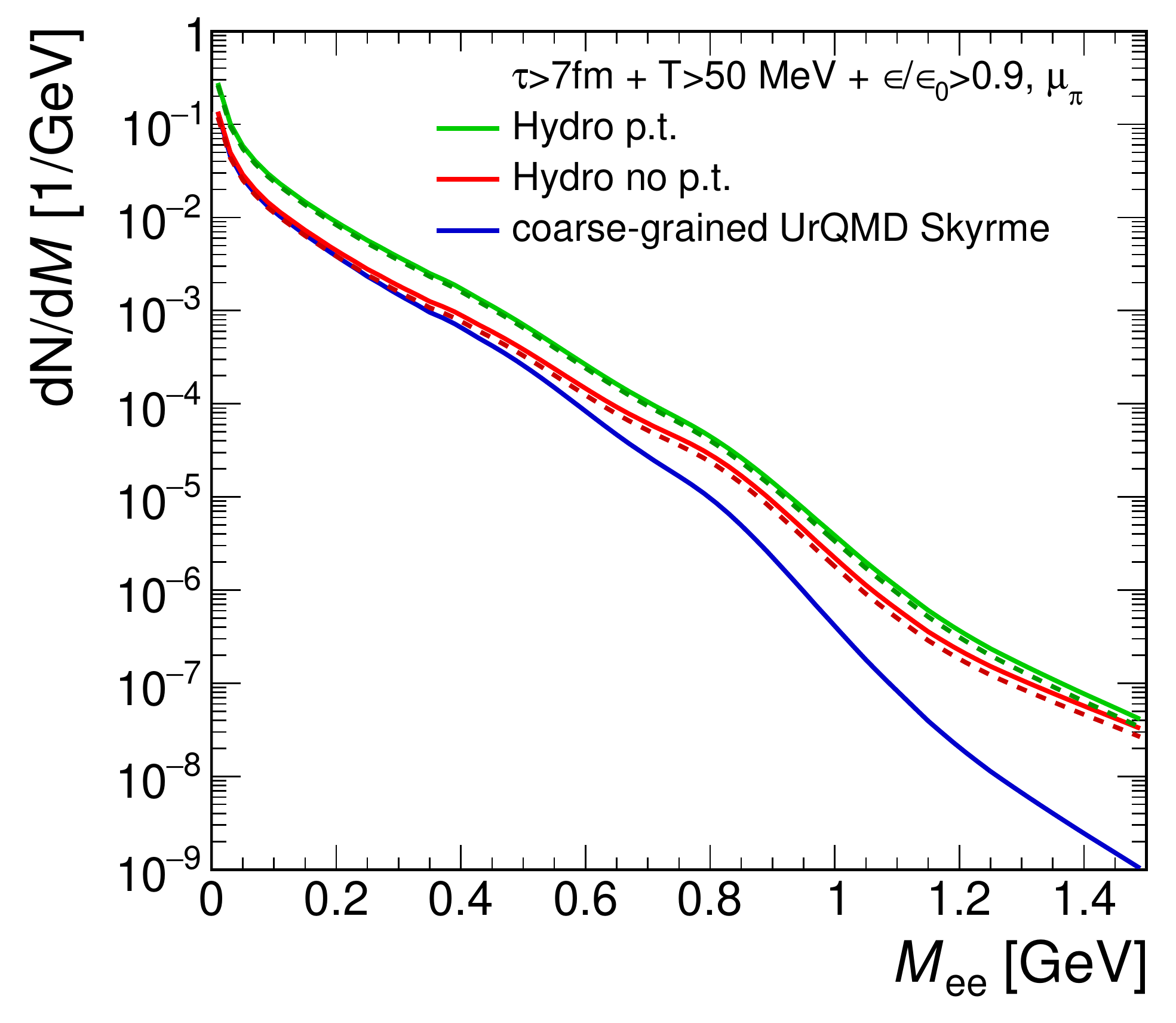}
   \hspace{0.5cm}
   \includegraphics[width=0.42\textwidth]{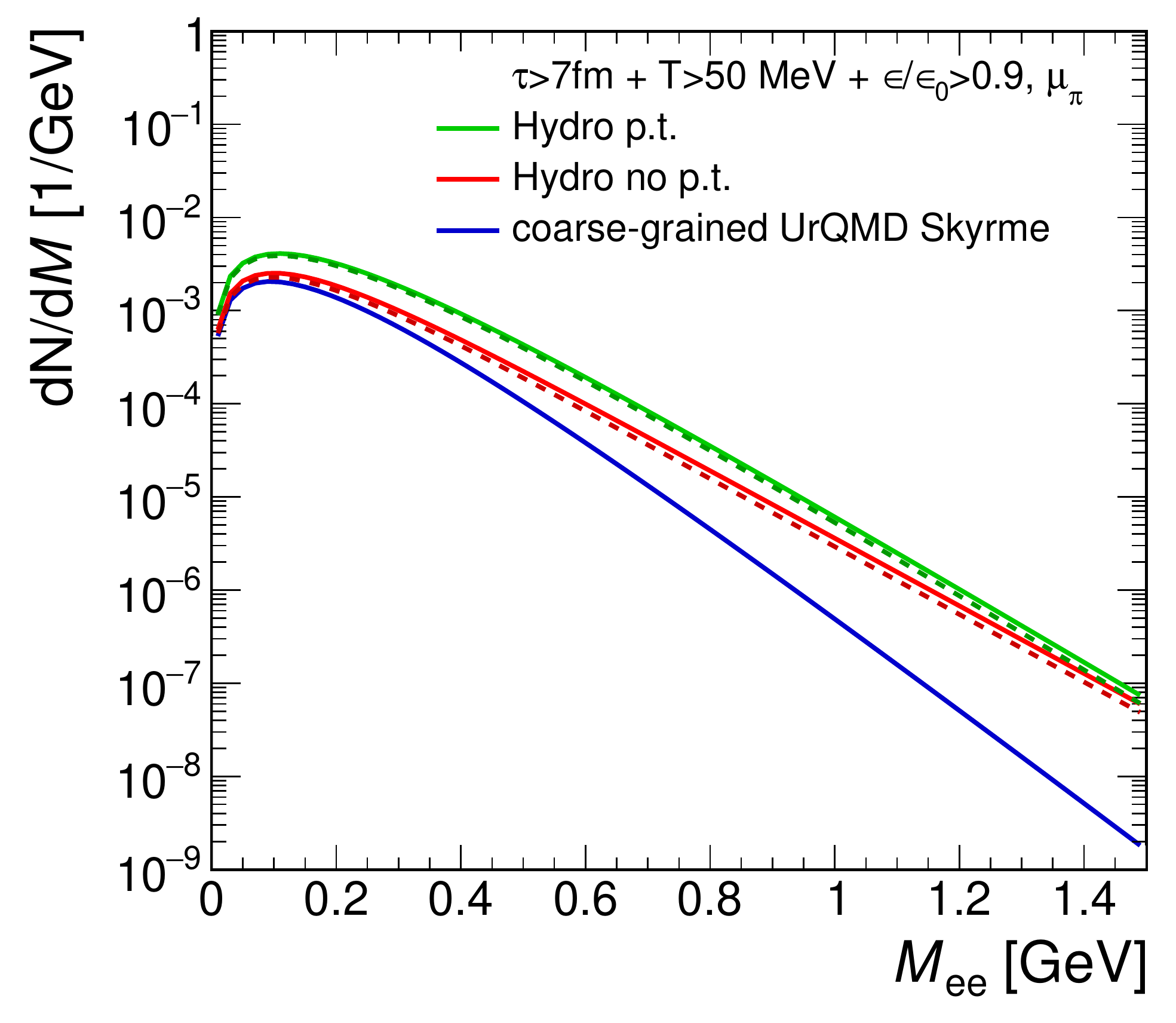}
   \caption{
   (Color online) 
   Invariant-mass spectrum of \ee\ radiated from central \auau\ collisions at 1.23 \agev.
   The blue, red and green curves show the results of the simulations with the coarse-grained UrQMD transport model, hydrodynamics with crossover EoS and with first-order EoS, respectively. The solid curves correspond to contributions from all cells while the dashed curves are calculated by accounting only for cells after 7 \fmc.
   The left and right panel correspond to results using the in-medium $\rho$-meson spectral function
   or the perturbative \qqbar\ rate, respectively.}
\label{fig:dilepton_spectra}
\end{figure}

Given our analysis of the space-time evolution of the thermal properties of the fireball in the previous section, we are now in a position to study their consequences on dilepton emission.
We convolute the mass- and momentum-dependent dilepton emission rates presented in Sec.~\ref{sec:emissivity} over discrete four-volumes with their local temperature and net-baryon density evolutions as input.
While the size of the four-volume elements in the fluid dynamical simulation, $0.2^3 \cdot 0.08 = 6.4 \times 10^{-4} {\rm fm}^4/c$, is substantially smaller than in the coarse-grained transport simulation, $1^3 \cdot 1 = 1 {\rm fm}^4/c$, we have checked that changing the latter does not create any noticeable difference in the results presented here.
We restrict the invariant-mass range to below 1.5 \gev\ in which the dilepton emission is dominated by the bulk of cells.
The emission spectra are evaluated on a kinematic grid that covers 1.5 units around midrapidity and $p_{\rm T}$ up to 4 \gev, \ie, practically full phase space.
The dilepton emission is then computed all the way down to thermal freeze-out, including the effects of chemical freeze-out to ensure pion-number conservation in both UrQMD and the crossover hydro simulation. 
In Fig.~\ref{fig:dilepton_spectra} we summarize the thermal dilepton invariant-mass spectra for the three different evolution models. In addition to the results using the in-medium hadronic rates in the left panel we also shown in the right panel spectra computed with the schematic leading-order perturbative rates. 

The strong broadening of the light vector mesons $\rho$ and $\omega$ induced by the hot and dense hadronic medium results in a remarkably structure-less radiation spectrum with only a slight bump structure (if any) remaining in the region of their vacuum masses for all three evolution models. However, the absolute yields differ substantially, especially between the first-order scenario and the no-PT evolutions. In the low-mass region (LMR) the first-order phase transition scenario clearly produces the largest yield, followed by the hydro with crossover EoS and the coarse-grained transport, which are quite comparable. More quantitatively, one finds the integrated yields in the mass range 0.3 to 0.7 \gev\  to be $N_{\rm {ee}}\times 10^4 = $ 4.1, 1.9 and 1.7 for the hydro version with phase-transition, without phase transition and the coarse-grained UrQMD, respectively.
The enhancement in the PT scenario can be attributed to the prolonged lifetime of the fireball due to the latent heat in the evolution (slowing down the expansion), as the low-mass radiation is known to be rather evenly distributed over the different evolution phases of the fireball~\cite{Rapp:2014hha}. This is the main result of our investigation. 

Toward higher masses, the spectrum of crossover hydro evolution features a significantly larger slope than the other two evolutions.
In comparison to the transport result this is not surprising, as the hydro achieves the maximal conversion of the incoming energy into thermal energy, but for the first-order case this is somewhat surprising, especially in view of the larger average temperatures of its cells, recall Fig.~\ref{fig:temp_dens_1d_7fm}.
Even the one-sigma spread does not suggest this. The inevitable conclusion is that the rather rare cells at the highest temperatures of near 110 \mev\ in the crossover evolution are the origin of this result (recall middle panel of Fig.~\ref{fig:3dhisto}).
More quantitatively, one can extract the pertinent slope parameters.
If the EM spectral function itself does not strongly depend on temperature (which is approximately the case for the in-medium broadened spectral functions), then the only temperature dependence of the dilepton emission rate is induced by the Boltzmann factor, cf.~Eq.~\ref{emisthermeq}). Then the invariant-mass spectrum is approximately proportional to $(MT)^{\frac{3}{2}} e^{-\frac{M}{T}}$.
Fitting this expression to the LMR of the spectra one finds $T_{\rm slope}=$ 86~\mev\ for the crossover hydro, 82~\mev\ for the hydro with phase-transition, and 75~\mev\ for coarse-grained UrQMD.
In the intermediate-mass range (IMR), for $M\ge 1$~\gev, temperatures of 119, 110, and 85~\mev\ are obtained from the three different models, respectively.
This corroborates the relevant (yet uncertain, as we will argue in the next section) role of the ``very hot" cells in the crossover evolution.

As a further step to distill the emission characteristics due to the different bulk scenarios, we have separated out the in-medium effects in the vector-meson spectral functions from the spectra 
by performing calculations with the perturbative \qqbar\ annihilation rate, Eq.~(\ref{rate_qqbar}), which is independent of $T$ and $\mu_{\rm B}$; the results are shown in the right panel of Fig.~\ref{fig:dilepton_spectra}.
We find that the essential features of the comparisons of the bulk scenarios found when using the in-medium spectral functions are preserved by these results.
This furthermore implies that the bulk dilepton emission is dominated by conditions where the broadening of the spectral functions is large, \ie, rather high baryon densities, since the perturbative \qqbar\ spectral function has no resonance structure.

\begin{figure}[tb]
\centering
   \includegraphics[width=0.42\textwidth]{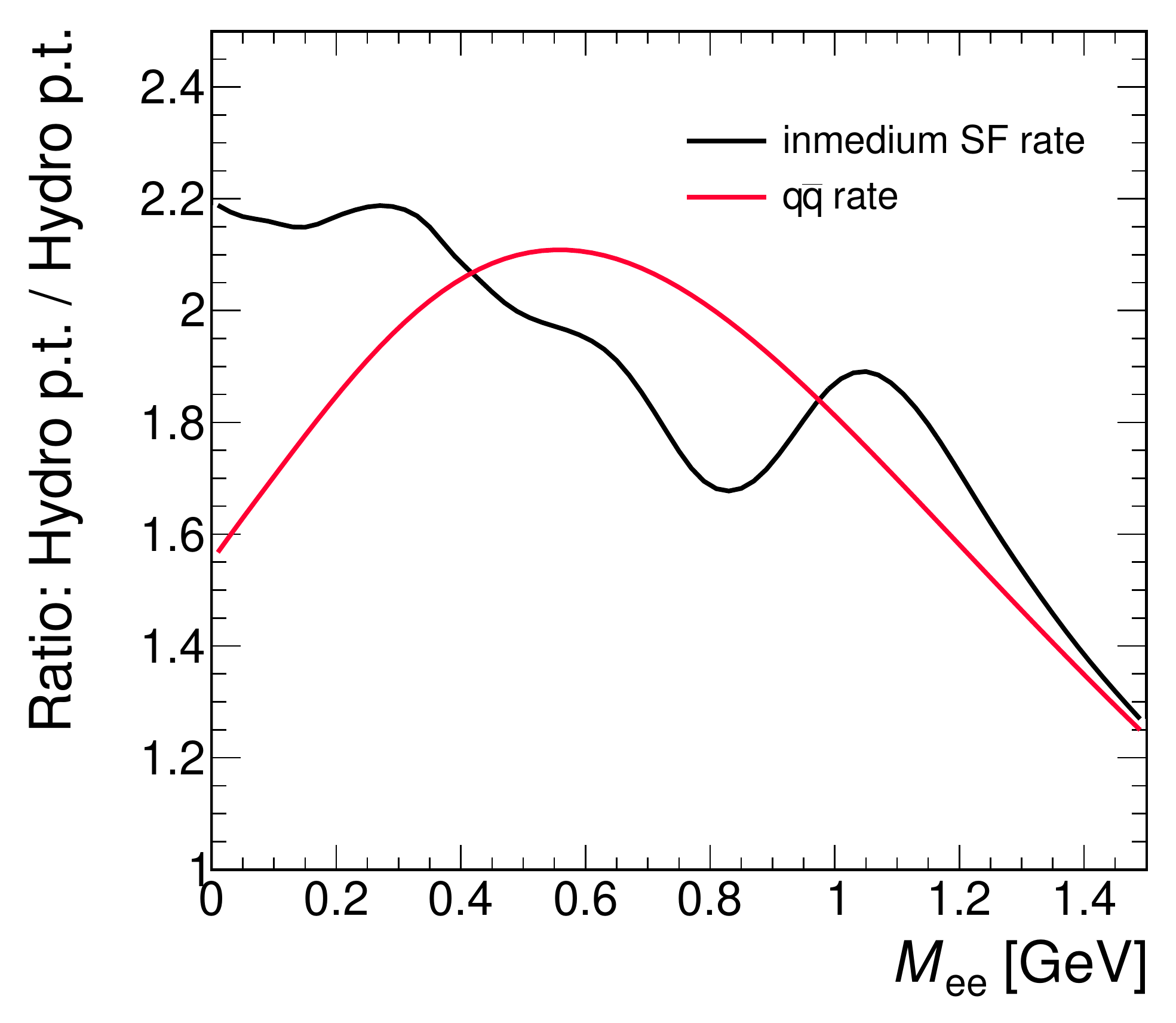}
   \hspace{0.5cm}
   \includegraphics[width=0.42\textwidth]{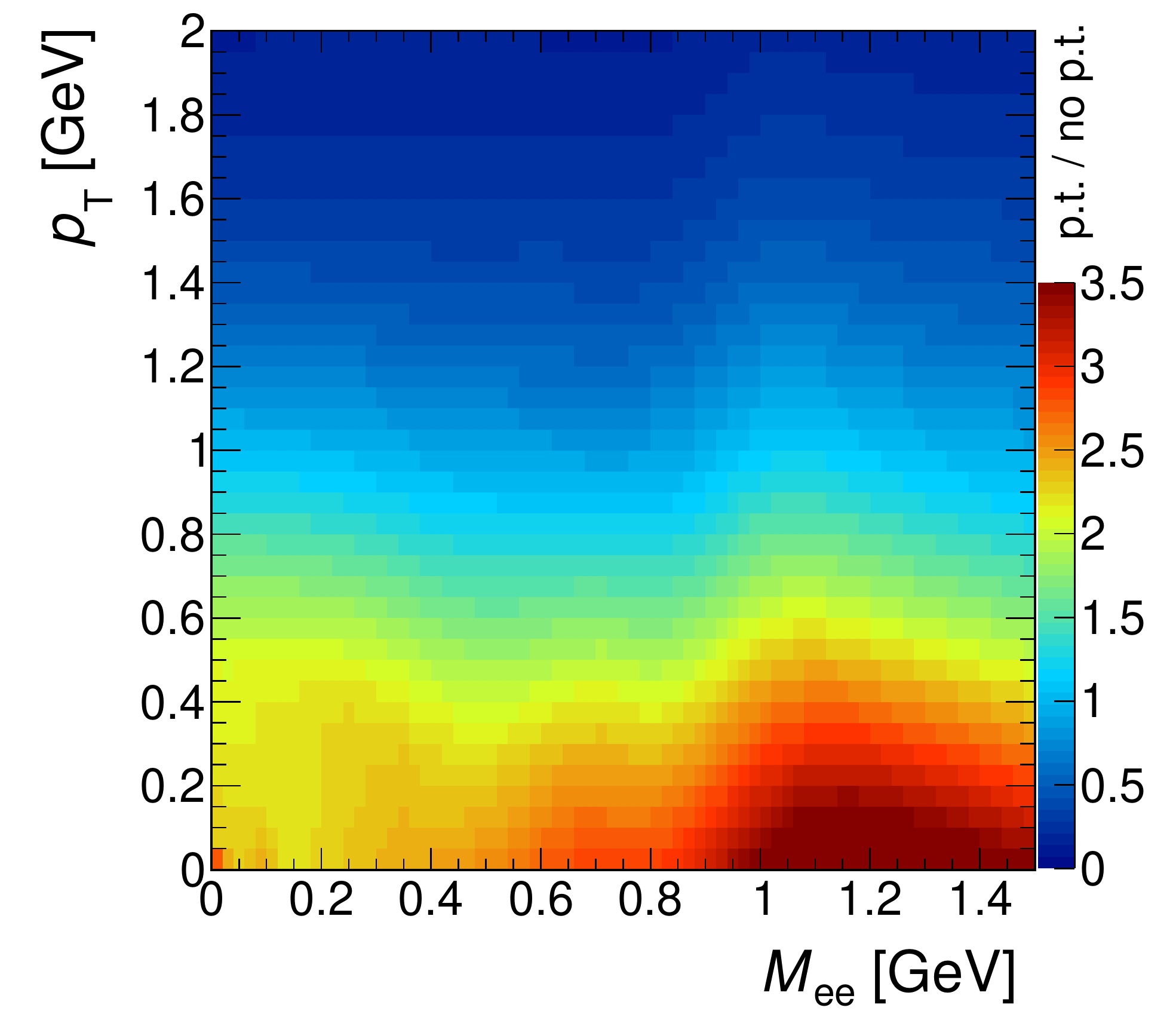}
   \caption{
   (Color online) 
   Left panel: Ratio of dilepton invariant-mass spectra from the hydro evolution with phase transition over the one without phase transition (here we only considered emission from cells after 7 \fmc). The result from using the in-medium vector-meson spectral functions (black line) is compared to the one using the \qqbar\ rate. 
   Right panel: Ratio of the 2-differential dilepton spectra in the plane of transverse pair momentum and invariant mass for the hydro with phase transition divided by the hydro without phase transition, using the in-medium hadronic emission rates. 
   }
\label{fig:dilepton_spectra_ratios}
\end{figure}
%


A more differential assessment of the enhancement effect due to the first-order phase transition is displayed in  Fig.~\ref{fig:dilepton_spectra_ratios} in terms of yield ratios between the first-order and crossover scenarios. The left panel displays the mass-differential (but still momentum-integrated) ratio of emitted dileptons. When using the emission rates based on the in-medium spectral functions the low-mass enhancement of the ratio of around a factor of 2 
is reflected  for invariant masses of around $M\simeq$ 0.4~\gev. Note that for the mass-integrated yields quoted above, the lower-mass region will have larger weights due to the near exponential rise in the absolute spectra.
The generally rising trend of the ratio at small $M$, and in particular the dip structure around the vacuum $\rho/\omega$ mass, are caused by the in-medium effects of the $\rho$ melting (low-mass enhancement and quenching of the peak), suggesting that in the first-order scenario the average densities probed in the emission are somewhat higher.
Again, to probe the robustness of our results with respect to the medium effects, we also plot the ratio for the case of the $\mu_{\rm B}$- and $T$-independent \qqbar\ rate.
As to be expected, the pertinent first-order to crossover ratio exhibits much less structure, but the overall enhancement of about a factor of 2 in the low-mass region persists, except at very small masses below $\approx$ 0.2~\gev\ where the \qqbar\ annihilation rate is lacking radiation processes.
With the in-medium hadronic rates, the broadening of the vector-meson spectral functions additionally enhances the ratio in the region at very small masses. This is caused by, \eg, Dalitz decays or Bremsstrahlung processes in the hadronic many-body calculation which are not present in the \qqbar\ rate  (which, \eg, gives a vanishing photon rate).
These sources are rather sensitive to the baryon density, reiterating that, on average, larger densities are probed by the radiation in the first-order scenario (recall the right panel of Fig.~\ref{fig:3dhisto}).
Thus, a quantitative control over the very-low mass region (both experimentally and theoretically) could in principle enhance the sensitivity to a first-order scenario.
The dropping trend of the mass ratio toward higher masses for the calculation with the \qqbar\ rate (and, above 1~\gev, also for the in-medium hadronic rate) suggests this feature to be due to a temperature effect, namely the larger spectral slope in the crossover scenario; its origin will be more closely inspected in the following section. 
In the right panel of Fig.~\ref{fig:dilepton_spectra_ratios} we show the 2-differential ratio as a contour plot in the plane of invariant mass and transverse pair momentum, $p_{\rm T}$. This plot indicates that the cuts focusing on small $p_{\rm T}$ can further augment the enhancement effect, even at relatively large mass (although largest factor of up to $\approx$~3 around $M\simeq 1.2$~\gev\ is subject to the same caveat mentioned above, which will be investigated in the following section).


\subsection{High-temperature cells and their impact on dilepton emission}
\label{ssec:highT}
As discussed earlier, the dilepton emission in the crossover scenario is characterized by a larger slope and almost catches up with the yields from the first-order scenario for dilepton masses near 1.5~\gev, despite a shorter lifetime and a smaller {\it average} temperature across the entire time evolution of the fireball (recall left panel of Fig.~\ref{fig:temp_dens_1d_7fm}). 
This suggests a substantial contribution from cells at very high temperatures, despite their small space-time weight, which we would like to further scrutinize in this section.

%
\begin{figure}[t]
\centering
   \includegraphics[width=0.8\textwidth]{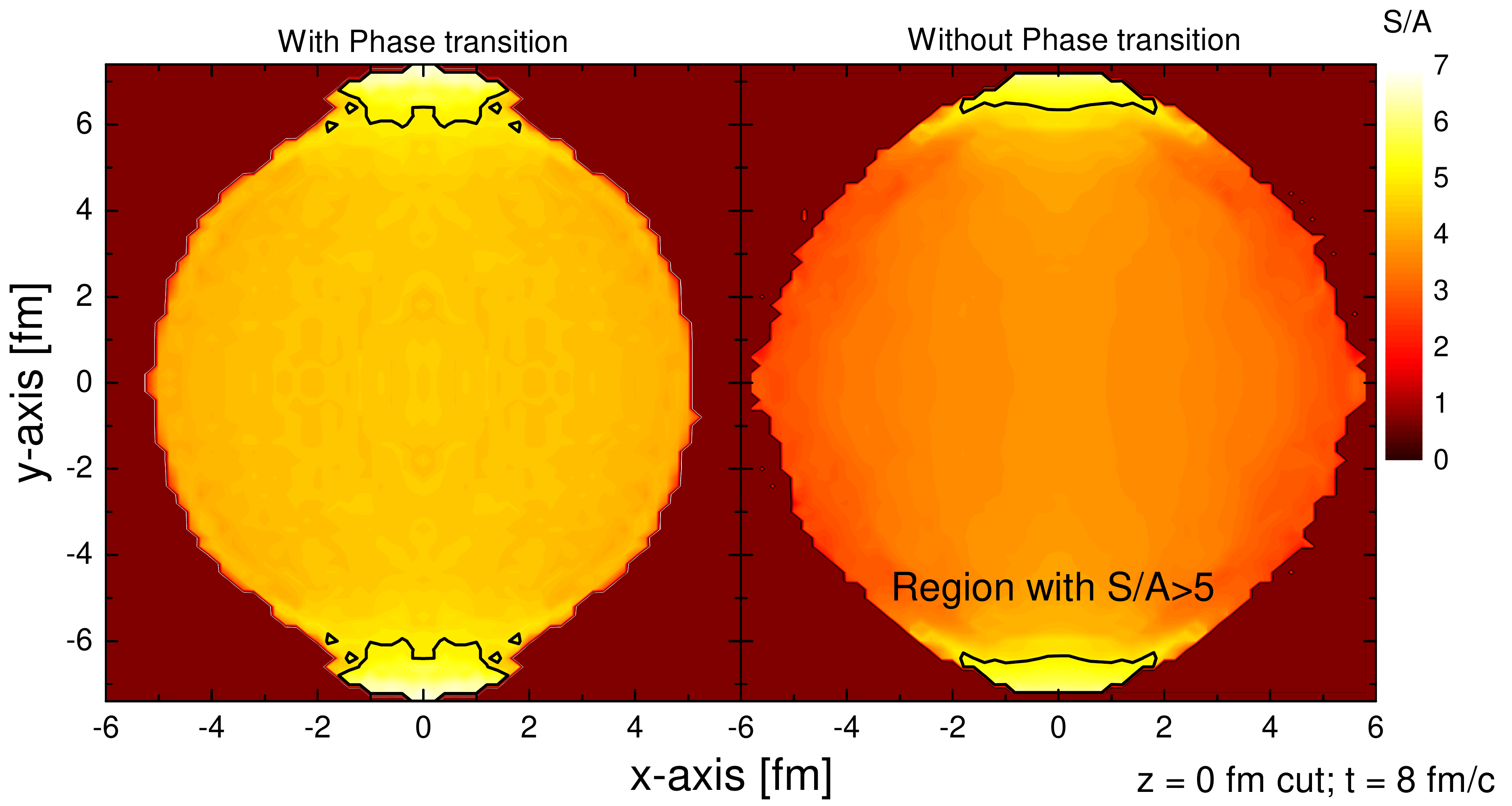}
   \caption{
   (Color online) Profile of the entropy per baryon, $S/$A, in the transverse plane ($x$-$y$ plane for $z=0$) in the fluid-dynamical  simulations with (left panel) and without (right panel) phase transition at a time of $t=8$ \fmc. Note the regions of maximal $S/$A at the upper and lower edges (``caps") of the hydrodynamic medium.
   }
\label{fig:hot_fluid_cells}
\end{figure}
\begin{figure}[t]
\centering
   \includegraphics[width=0.42\textwidth]{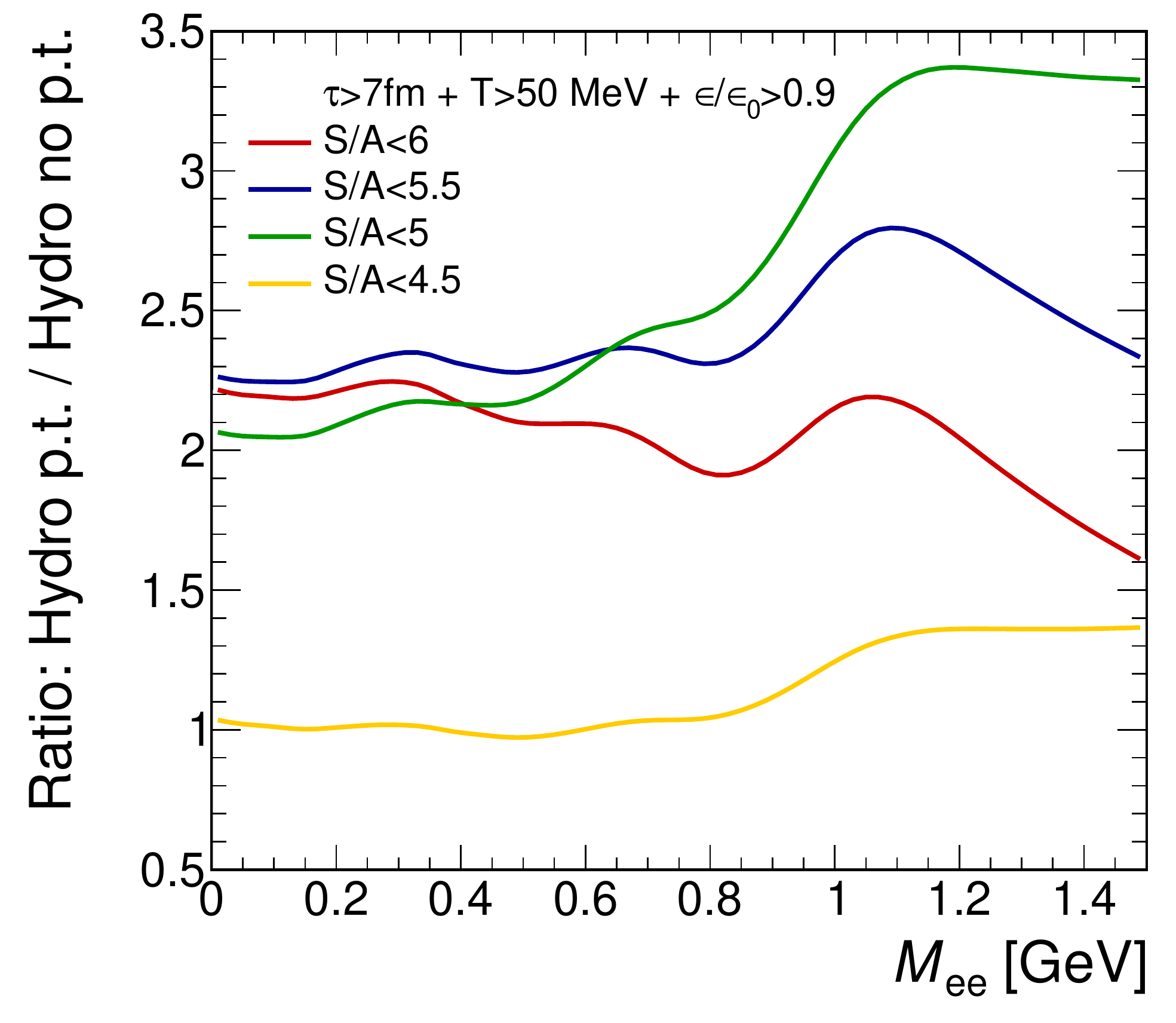}
   \hspace{0.5cm}
   \includegraphics[width=0.42\textwidth]{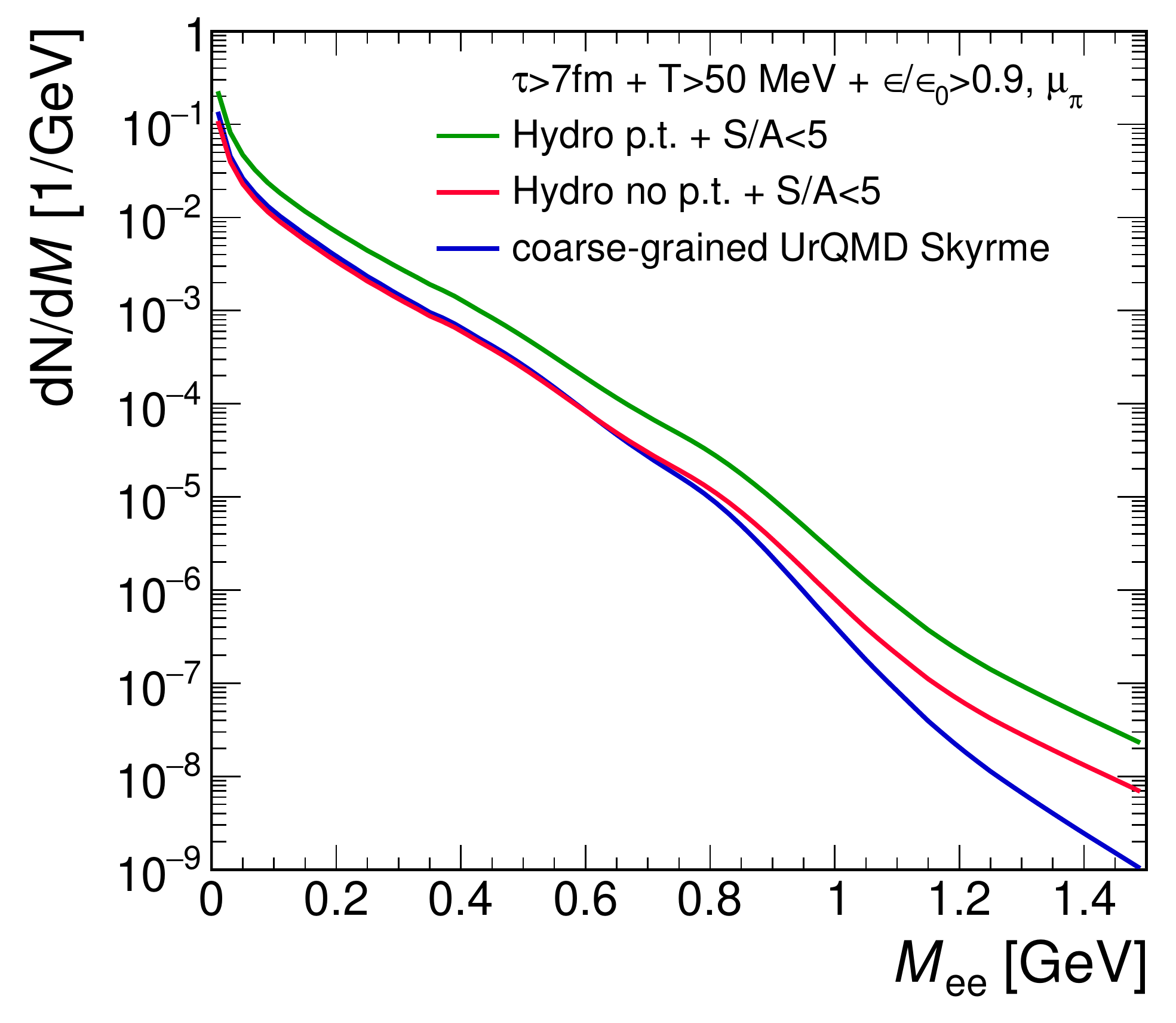}
   \caption{
   (Color online) Left panel: Ratio of dilepton spectra from the hydro evolution with phase transition divided by the one without, for different $S/$A cuts. Only cells from times 
   $>7$ \fmc\ are taken into account, and the in-medium vector-meson spectral functions are used.
   Right panel: Comparison of the \ee\ invariant-mass spectra from the hydro evolutions with $S/{\rm A} < 5$ (green and red line with and without phase transition, respectively) to the coarse-grained UrQMD transport simulation (blue line).}
\label{fig:mass_SoverA_cuts}
\end{figure}

Toward this end we display in Fig.~\ref{fig:hot_fluid_cells} the distribution of the entropy per baryon, $S/$A, in all fluid cells in the transverse plane of a central 1.23~\agev\ \auau\ collision at time $t=8$ \fmc\ for $z=0$, which is in the vicinity of the formation of the maximal energy densities in the collision.
One clearly identifies two regions of maximal $S/$A of up to 6 to 7, essentially the upper and lower ``caps" of the reaction zone, which correspond to the hottest spots with temperatures larger than $T>105$ \mev.
One may understand the occurrence of these hot regions as a result of the so-called squeeze-out effect.
As the two incoming nuclei make contact, the overlap regions at the top and bottom rapidly build up significant flow into the direction of the vacuum (facilitated by the instant-thermalization assumption), the out-of-plane direction. Contrary to expansion into the direction of the reaction plane, where the matter encounters relatively cold nuclear matter which tames the entropy per baryon, this is not the case for the out-of plane direction. Together with the rapid squeeze out, the entropy per baryon rises in this direction leading to relatively large temperatures. The size of this region is, however, only a few fm$^3$.
In the coarse grained UrQMD simulation, which also contains the squeeze-out effect, a similar entropy production is not observed. It is therefore quite possible that this effect is an artefact of the instant thermalization assumption in the ideal-hydro description, especially since it occurs on the fringes of the fireball where the validity of this assumption is diminished.
To quantify the contributions of the hot caps to the dilepton spectra, we display in Fig.~\ref{fig:mass_SoverA_cuts} (left panel) the ratio of the spectra with phase transition over those without when imposing varying upper cutoffs on the $S/$A ratio, starting with 6 and then lowered in steps of 0.5. With $(S/{\rm A})_{\rm max} = 6$, the results for $M\lsim0.6$ \gev\ agree within 10\% with the ones without cutoff, a trend that persists when subsequently lowering $(S/{\rm A})_{\rm max}$ down to 5. However, for $(S/{\rm A})_{\rm max} = 4.5$, a major drop of the ratio by a factor of $\approx 2$ is observed, signaling that this cut eliminates about half of the cells of the bulk dilepton emission for the phase transition scenario, while only a small fraction of cells is removed in case of the hydro without phase transition.
At higher invariant masses, cutoff values for $S/$A larger than 4.5 have a much more noticeable impact than at low mass: the falling trend of this ratio without cut is appreciably mitigated, even reversed down to $(S/{\rm A})_{\rm max}\simeq5$, reiterating the stronger role of the high-temperature cells at higher mass. However, when lowering $(S/{\rm A})_{\rm max}$ to 4.5, also the high-mass ratio abruptly drops down close to one.
%


In the right-hand panel of Fig.~\ref{fig:mass_SoverA_cuts} we compare the invariant-mass spectra of both hydro results with the $S/{\rm A}<5$ cut with the UrQMD coarse-grained results. 
The transport and hydro-crossover results are now very similar, differing by less than $\approx$10\% in the LMR.
This confirms that the main difference between the hydro-crossover evolution and the coarse-grained transport lies in the high temperature tail. The enhancement of the dilepton yield in the LMR for the first-order scenario can therefore be considered a robust result as being due to the mixed phase of the fluid dynamical simulation. 

\section{Conclusions}
\label{sec:concl}
In this work we have investigated the impact of different evolution 
models on the properties of thermal dilepton spectra in Au-Au collisions at HADES energies. In particular, we have compared the results for two different hydrodynamic simulations, employing an EoS with and without a first-order phase transition, and a coarse-grained transport simulation based on the UrQMD model with mean-field effects. 
All of these evolutions have been benchmarked against the final-state pion spectra, showing good agreement. 
The thermal dilepton rates have been taken from existing calculations of in-medium vector-meson spectral functions.
Our main result is that a first-order phase transition leads to a substantial increase of the low-mass thermal dilepton yield over that from a crossover transition, by about a factor two, as a consequence of the prolonged lifetime caused by the mixed-phase formation. 
At the same time, the dilepton spectra from the crossover evolution show good agreement with the ones from coarse-grained transport, especially when removing hot spots at the fringes of the fireball in the hydro evolutions which we have attributed to an artefact of the instant-thermalization assumption. We have verified that removing the hot spots from the hydrodynamics does not affect the enhancement due to the mixed-phase formation in the first-order scenario. We also found that the in-medium effects in the spectral functions, \ie, the strong broadening of the vector-meson peaks, lead to an additional relative enhancement at masses around 0.2~\gev\ in the first-order scenario, reflecting the higher average densities produced in the more compressible medium with mixed phase.

Our findings call for further quantitative investigations in comparison to existing HADES data. This work is in progress. Also, based on the notion that the first-order transition should disappear at lower chemical potentials and higher temperatures, an experimental measurement of the excitation function in the the SIS100 beam energy range appears to 
be a promising undertaking.

\begin{acknowledgments}
J.S. and A.M. thank Stefan Schramm for helpful discussions and guidance on the CMF model. This work was supported by the Deutsche Forschungsgemeinschaft (DFG) through the grant No. CRC-TR211 ``Strong-interaction matter under extreme conditions,” by the U.S. National Science Foundation (NSF) under grant No.~PHY-1913286 and the ExtreMe Matter Institute (R.R.), by the Hungarian K10103/18-D0200 kiv\'al\'os\'agi program NKFIH FK-123842 (A.M.), and by the Samson AG and the BMBF through the ErUM-Data project (J.S.).
\end{acknowledgments}

\bibliography{bibnew}
\end{document}